\documentstyle[twoside,epsfig,psfig]{article}

\catcode`\@=11
\long\def\@makefntext#1{
\protect\noindent \hbox to 3.2pt {\hskip-.9pt  
$^{{\eightrm\@thefnmark}}$\hfil}#1\hfill}		

\def\@makefnmark{\hbox to 0pt{$^{\@thefnmark}$\hss}}	
	
\def\ps@myheadings{\let\@mkboth\@gobbletwo
\def\@oddhead{\hbox{}
\rightmark\hfil\eightrm\thepage}   
\def\@oddfoot{}\def\@evenhead{\eightrm\thepage\hfil
\leftmark\hbox{}}\def\@evenfoot{}
\def\sectionmark##1{}\def\subsectionmark##1{}}

\oddsidemargin=\evensidemargin
\newcounter{sectionc}\newcounter{subsectionc}\newcounter{subsubsectionc}
\renewcommand{\section}[1] {\vspace{12pt}\addtocounter{sectionc}{1} 
\setcounter{subsectionc}{0}\setcounter{subsubsectionc}{0}\noindent 
	{\tenbf\thesectionc. #1}\par\vspace{5pt}}
\renewcommand{\subsection}[1] {\vspace{12pt}\addtocounter{subsectionc}{1} 
	\setcounter{subsubsectionc}{0}\noindent 
	{\bf\thesectionc.\thesubsectionc. {\kern1pt \bfit #1}}\par\vspace{5pt}}
\renewcommand{\subsubsection}[1] {\vspace{12pt}\addtocounter{subsubsectionc}{1}
	\noindent{\tenrm\thesectionc.\thesubsectionc.\thesubsubsectionc.
	{\kern1pt \tenit #1}}\par\vspace{5pt}}

\newcounter{appendixc}
\newcounter{subappendixc}[appendixc]
\newcounter{subsubappendixc}[subappendixc]
\renewcommand{\thesubappendixc}{\Alph{appendixc}.\arabic{subappendixc}}
\renewcommand{\thesubsubappendixc}
	{\Alph{appendixc}.\arabic{subappendixc}.\arabic{subsubappendixc}}

\renewcommand{\appendix}[1] {\vspace{12pt}
        \refstepcounter{appendixc}
        \setcounter{figure}{0}
        \setcounter{table}{0}
        \setcounter{lemma}{0}
        \setcounter{theorem}{0}
        \setcounter{corollary}{0}
        \setcounter{definition}{0}
        \setcounter{equation}{0}
        \renewcommand{\thefigure}{\Alph{appendixc}.\arabic{figure}}
        \renewcommand{\thetable}{\Alph{appendixc}.\arabic{table}}
        \renewcommand{\theappendixc}{\Alph{appendixc}}
        \renewcommand{\thelemma}{\Alph{appendixc}.\arabic{lemma}}
        \renewcommand{\thetheorem}{\Alph{appendixc}.\arabic{theorem}}
        \renewcommand{\thedefinition}{\Alph{appendixc}.\arabic{definition}}
        \renewcommand{\thecorollary}{\Alph{appendixc}.\arabic{corollary}}
        \renewcommand{\theequation}{\Alph{appendixc}.\arabic{equation}}
        \noindent{\tenbf Appendix \theappendixc #1}\par\vspace{5pt}}
\newcommand{\subappendix}[1] {\vspace{12pt}
        \refstepcounter{subappendixc}
        \noindent{\bf Appendix \thesubappendixc. {\kern1pt \bfit #1}}
	\par\vspace{5pt}}
\newcommand{\subsubappendix}[1] {\vspace{12pt}
        \refstepcounter{subsubappendixc}
        \noindent{\rm Appendix \thesubsubappendixc. {\kern1pt \tenit #1}}
	\par\vspace{5pt}}

\newcommand{\textlineskip}{\baselineskip=13pt}
\newcommand{\smalllineskip}{\baselineskip=10pt}

\def\eightcirc{
\begin{picture}(0,0)
\put(4.4,1.8){\circle{6.5}}
\end{picture}}
\def\eightcopyright{\eightcirc\kern2.7pt\hbox{\eightrm c}} 

\newcommand{\copyrightheading}[1]
	{\vspace*{-2.5cm}\smalllineskip{\flushleft
	{\footnotesize International Journal of Modern Physics E, #1}\\
	{\footnotesize $\eightcopyright$\, World Scientific Publishing
	 Company}\\
	 }}

\newcommand{\publisher}[2]{{\begin{center}\footnotesize\smalllineskip 
	Received #1\\
	Revised #2
	\end{center}
	}}

\def\abstracts#1#2#3{{
	\centering{\begin{minipage}{4.5in}\baselineskip=10pt\footnotesize
	\parindent=0pt #1\par 
	\parindent=15pt #2\par
	\parindent=15pt #3
	\end{minipage}}\par}} 

\renewenvironment{thebibliography}[1]
	{\frenchspacing
	 \ninerm\baselineskip=11pt
	 \begin{list}{\arabic{enumi}.}
        {\usecounter{enumi}\setlength{\parsep}{0pt}     
	 \setlength{\leftmargin 12.7pt}{\rightmargin 0pt} 
         \setlength{\itemsep}{0pt} \settowidth
	{\labelwidth}{#1.}\sloppy}}{\end{list}}

\newcounter{itemlistc}
\newcounter{romanlistc}
\newcounter{alphlistc}
\newcounter{arabiclistc}

\newcommand{\fcaption}[1]{
        \refstepcounter{figure}
        \setbox\@tempboxa = \hbox{\footnotesize Fig.~\thefigure. #1}
        \ifdim \wd\@tempboxa > 5in
           {\begin{center}
        \parbox{5in}{\footnotesize\smalllineskip Fig.~\thefigure. #1}
            \end{center}}
        \else
             {\begin{center}
             {\footnotesize Fig.~\thefigure. #1}
              \end{center}}
        \fi}

\newcommand{\tcaption}[1]{
        \refstepcounter{table}
        \setbox\@tempboxa = \hbox{\footnotesize Table~\thetable. #1}
        \ifdim \wd\@tempboxa > 5in
           {\begin{center}
        \parbox{5in}{\footnotesize\smalllineskip Table~\thetable. #1}
            \end{center}}
        \else
             {\begin{center}
             {\footnotesize Table~\thetable. #1}
              \end{center}}
        \fi}

\def\@citex[#1]#2{\if@filesw\immediate\write\@auxout
	{\string\citation{#2}}\fi
\def\@citea{}\@cite{\@for\@citeb:=#2\do
	{\@citea\def\@citea{,}\@ifundefined
	{b@\@citeb}{{\bf ?}\@warning
	{Citation `\@citeb' on page \thepage \space undefined}}
	{\csname b@\@citeb\endcsname}}}{#1}}

\newif\if@cghi
\def\cite{\@cghitrue\@ifnextchar [{\@tempswatrue
	\@citex}{\@tempswafalse\@citex[]}}
\def\citelow{\@cghifalse\@ifnextchar [{\@tempswatrue
	\@citex}{\@tempswafalse\@citex[]}}
\def\@cite#1#2{{$\null^{#1}$\if@tempswa\typeout
	{IJCGA warning: optional citation argument 
	ignored: `#2'} \fi}}

\def\pmb#1{\setbox0=\hbox{#1}
	\kern-.025em\copy0\kern-\wd0
	\kern.05em\copy0\kern-\wd0
	\kern-.025em\raise.0433em\box0}

\def\fnt#1#2{\footnotetext{\kern-.3em
	{$^{\mbox{\scriptsize #1}}$}{#2}}}

\def\fpage#1{\begingroup
\voffset=.3in
\thispagestyle{empty}\begin{table}[b]\centerline{\footnotesize #1}
	\end{table}\endgroup}

\def\runninghead#1#2{\pagestyle{myheadings}
\markboth{{\protect\footnotesize\it{\quad #1}}\hfill}
{\hfill{\protect\footnotesize\it{#2\quad}}}}
\font\tenrm=cmr10
\font\tenit=cmti10 
\font\tenbf=cmbx10
\font\bfit=cmbxti10 at 10pt
\font\ninerm=cmr9

\font\eightrm=cmr8

\def\qed{\hbox{${\vcenter{\vbox{			
   \hrule height 0.4pt\hbox{\vrule width 0.4pt height 6pt
   \kern5pt\vrule width 0.4pt}\hrule height 0.4pt}}}$}}

\def\bsc{{\sc a\kern-6.4pt\sc a\kern-6.4pt\sc a}}	
\def\bflatex{\bf L\kern-.30em\raise.3ex\hbox{\bsc}\kern-.14em 
T\kern-.1667em\lower.7ex\hbox{E}\kern-.125em X} 

\begin{document}

\runninghead{Hyperon Interactions $\ldots$} {Hyperon Interactions $\ldots$}

\normalsize\textlineskip
\thispagestyle{empty}
\setcounter{page}{1}

\copyrightheading{}			

\vspace*{0.88truein}

\fpage{1}
\centerline{\bf WEAK AND ELECTROMAGNETIC INTERACTIONS OF HYPERONS:}
\vspace*{0.035truein}
\centerline{\bf A CHIRAL APPRACH}
\vspace*{0.37truein}
\centerline{\footnotesize BARRY R. HOLSTEIN}
\vspace*{0.015truein}
\centerline{\footnotesize\it Institute for Nuclear Theory, Department
of Physics, University
of Washington}
\baselineskip=10pt
\centerline{\footnotesize\it Seattle, WA  98195}
\vspace*{10pt}
\centerline{\footnotesize and}
\vspace*{10pt}
\centerline{\footnotesize\it Department of Physics, 
University of Massachusetts}
\baselineskip=10pt
\centerline{\footnotesize\it Amherst, MA  01003}
\vspace*{0.225truein}
\publisher{(received date)}{(revised date)}

\vspace*{0.21truein}
\abstracts{A range of issues in the field of weak and electromagnetic interactions of
hyperons is presented from the perspective of (broken) chiral symmetry, together
with an assessment of where important challenges remain.}{}{}


\vspace*{1pt}\textlineskip	
\section{Introduction}		
\vspace*{-0.5pt}
\noindent
 
In their 1947 study of cosmic rays Rochester and Butler obtained 
evidence for the existence of massive
unstable particles\cite{rb}.  While examining
muon and electron showers using a cloud chamber they secured photographs
clearly demonstrating the decay of a neutral system into two charged particles
as well as a charged system into a charged plus neutral particle.  The authors
were able to assign lower limits of $770\pm 200m_e$ and $980\pm 150m_e$ to
the masses of these neutral and charged particles respectively.  We now 
recognize this as being the discovery of the hyperons.  During the intervening
half century much has been learned about the properties of such particles, but 
many basic questions still remain, and we shall try in this article 
to summarize the state of the field from a chiral symmetry perspective, 
at least as far as electromagnetic and weak hyperon interactions are
concerned.   During the very early days, hyperon studies focused on  
basic particle properties.   Then during the 1960's a deeper 
understanding was developed in terms of a three-quark picture and elementary 
symmetries---SU(3) in
particular.  During the late 1980's and 1990's the general approach shifted
once again and is now generally based upon (broken) chiral symmetry, 
which is a fundamental property of quantum chromodynamics (QCD) 
believed to underlie 
all of particle/nuclear interactions.  However, since the QCD Lagrangian 
is written 
in terms of quark/gluon degrees of freedom, it is not {\it directly} applicable
to low energy hadronic physics.  Rather one exploits QCD at low energies via 
its (chiral) symmetry properties and applies it using 
the technique of effective 
interactions.  Since such methods have become such a fixture in contemporary
discussions but are not yet generally familiar, we begin our own
presentation with an introduction to chiral perturbation theory and effective
field theory, before 
moving to the subject of hyperons and their weak and electromagnetic 
interactions, which is the main focus of our report.  Note that, for space
reasons and in order to maintain a coherent focus, we will
omit an important area of interest in the physics of hyperons: strong
interactions and polarization issues in hyperon production\cite{rad}.       

\section{Chiral Symmetry: a Brief Introduction}

In the early days of the field (say forty years ago) the holy grail 
of particle/nuclear physicists was to construct a theory of elementary 
particle interactions which emulated quantum electrodynamics in that it 
was elegant, renormalizable, and phenomenological successful.  It is now 
four decades later and we have identified a theory which satisfies two 
out of the three criteria---quantum chromodynamics (QCD).
Indeed the form of the QCD Lagrangian\footnote{Here the covariant derivative is
\begin{equation}
i D_{\mu}=i\partial_{\mu}-gA_\mu^a {\lambda^a \over 2} \, ,
\end{equation}
where $\lambda^a$ (with $a=1,\ldots,8$) are the SU(3) Gell-Mann matrices,
operating in color space, and the color-field tensor is defined by
\begin{equation}
G_{\mu\nu}=\partial_\mu  A_\nu -  \partial_\nu  A_\mu -
g [A_\mu,A_\nu]  \, ,
\end{equation} }
\begin{equation}
{\cal L}_{\mbox{\tiny QCD}}=\bar{q}(i  {\not\!\! D} - m )q-
{1\over 2} {\rm tr} \; G_{\mu\nu}G^{\mu\nu} \, .
\end{equation}
is elegantly simple, and the theory is renormalizable.  So why are 
physicists still not
satisfied?  The difficulty lies with the third criterion---phenomenological 
success.  While at the very largest energies, asymptotic freedom allows the 
use of perturbative
techniques, for those who are interested in making contact with low energy 
experimental findings, such as will be attempted in this paper, 
there exist at least three fundamental difficulties:
\begin{itemize}
\item [i)] QCD is written in terms of the "wrong" degrees of 
freedom---quarks and
gluons---while experiments are performed with hadronic bound states;

\item [ii)] the theory is hopelessly non-linear due to gluonic 
self interaction;

\item[iii)] the theory is one of strong coupling---$g^2/4\pi\sim 1$---so that 
perturbative methods are not practical.
\end{itemize}
Nevertheless, there has been a great deal of recent progress in making 
contact between
theory and experiment using the technique of "effective 
field theory."\cite{eft}

In order to obatain a feel for this idea, one must realize that there are 
many situations in physics which involve two very different mass 
scales---one heavy and one light.
Then, provided that one is working with energy-momenta small 
compared to the heavy
scale, one can treat the heavy degrees of freedom purely in terms of 
their virtual effects.  Indeed by the uncertainty principle, such effects 
can be included in simple short distance---local---interactions.  
Effective field theory then describes the low energy ramifications of 
physics arising from large energy scales in terms of parameters in a 
local Lagrangian, which can be measured phenomenologically or calculated 
from a more complete theory.  The low energy component of the theory is 
not subject to this simplification---it is
non-local and must be treated fully quantum mechanically.  Such theories are
nonrenormalizable, but so be it.  They are calculable and allow reliable 
predictions to be made for experimental observables.

In the case of QCD, what makes the effective field theory---called chiral 
perturbation theory\cite{cpt},\cite{dsm}---appropriate is the feature, to
be discussed below, that symmetry under SU(3) axial transformations is 
spontaneously broken, leading to the presence of light so-called Goldstone 
bosons---$\pi,K,\eta$---which must couple to one another and to other 
particles {\it derivatively}.  This implies that {\it low energy} Goldstone 
interactions are weak and may be treated perturbatively.  The
light energy scale in this case is set by the Goldstone masses---several 
hundred MeV---while the heavy scale is everything else---$M_N,m_\rho,{\it etc.}
\sim 1$ GeV, so that we expect our effective field theory to be 
"effective" provided that $E,|\vec{p}|<<1$ GeV.
         
Before becoming more explicit about application of effective interaction
ideas within a quantum field theoretic context, however, it is useful to 
cite a familiar example from the realm of ordinary quantum 
mechanics---Rayleigh scattering.
         
\subsection{Rayleigh Scattering}

Before proceeding to QCD, we study effective field theory
in the simpler context of ordinary quantum mechanics, in order to get familiar 
with the idea.
Specifically, we look at the question of why the sky is blue, whose 
answer can be
found in an analysis of the scattering of photons from the sun 
by atoms in the atmosphere---Compton scattering.\cite{skb}  
First we examine the 
problem using traditional quantum mechanics and, for simplicity, consider 
elastic (Rayleigh) scattering from
single-electron (hydrogen) atoms.  The appropriate Hamiltonian is then
\begin{equation}
H={(\vec{p}-e\vec{A})^2\over 2m}+e\phi
\end{equation}
and the leading---${\cal O}(e^2)$---amplitude for Compton scattering
is given by the Kramers-Heisenberg formula, which arises from the 
Feynman diagrams shown in Figure 1---
\begin{eqnarray}
{\rm Amp}&=&-{e^2/m\over \sqrt{2\omega_i2\omega_f}}\left[\hat{\epsilon}_i\cdot
\hat{\epsilon}_f^*+{1\over m}\sum_n\left({\hat{\epsilon}_f^*\cdot
<0|\vec{p}e^{-i\vec{q}_f\cdot\vec{r}}|n>
\hat{\epsilon}_i\cdot <n|\vec{p}e^{i\vec{q}_i\cdot\vec{r}}|0>\over 
\omega_i+E_0-E_n}\right.\right.
\nonumber\\
&+&\left.\left.{\hat{\epsilon}_i\cdot <0|\vec{p}e^{i\vec{q}_i\cdot\vec{r}}|n>
\hat{\epsilon}_f^*\cdot <n|\vec{p}e^{-i\vec{q}_f\cdot\vec{r}}|0>\over E_0-\omega_f-E_n}
\right)\right]
\end{eqnarray}
where $|0>$ represents the hydrogen ground state having binding energy
$E_0$. 

\begin{figure}
\begin{center}
\epsfig{file=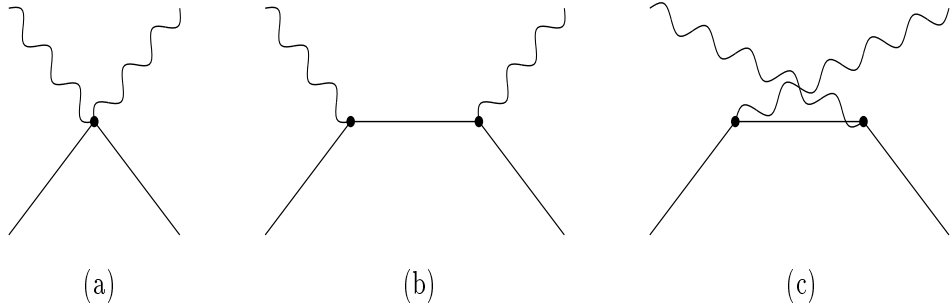,height=4cm,width=12.5cm}
\caption{Feynman diagrams contributing to low energy Compton scattering.}
\end{center}
\end{figure}

\noindent (Note that for simplicity 
we take the proton to be infinitely heavy so
it need not be considered.)  Here the leading component is the familiar 
$\omega$-independent Thomson amplitude and would appear naively 
to lead to an energy-independent
cross-section.  However, this is {\it not} the case.  Indeed, provided that the
energy of the photon is smaller than a typical excitation energy (as is the
case for optical photons), it can be 
shown by expanding in powers of $\omega/\Delta E$, that the cross section 
can be written as\cite{bh},\cite{sak}
\begin{eqnarray}
\quad {d\sigma\over d\Omega}&=&
\lambda^2\omega^4|\hat{\epsilon}_{f}^*\cdot
\hat{\epsilon}_{i}|^2\left(1+{\cal O}\left({\omega^2\over (\Delta E)^2}
\right)\right)\label{eq:cc}
\end{eqnarray}
where 
\begin{equation}
\lambda=\alpha_{em}\sum_n{2|z_{n0}|^2\over E_n-E_0}
\end{equation}
is the atomic electric polarizability,
$\alpha_{em}=e^2/4\pi$ is the fine structure constant, and $\Delta E\sim m
\alpha_{em}^2$ is a typical hydrogen excitation energy.  We note that  
$\lambda\sim a_0^2\times {\alpha_{em}\over \Delta E}\sim a_0^3$ 
is of order the atomic volume, as will be exploited below and that the cross 
section itself has the characteristic $\omega^4$ 
dependence which leads to the blueness of the sky---blue light scatters much
more strongly than red.\cite{feyn}

Now while the above derivation is rigorous and correct, 
it requires somewhat detailed and
lengthy quantum mechanical manipulations, which obscure the relatively 
simple physics
involved.  One can avoid these problems by the use of effective field theory 
methods outlined above.  The key point is that of scale.  Since the 
incident photons
have wavelengths $\lambda\sim 5000$A much larger than the $\sim$ 1A 
atomic size, then at 
leading order the photon is insensitive to the presence of the atom, since 
the latter is electrically neutral---the effective leading order Hamiltonian 
is simply that for the hydrogen atom 
\begin{equation}
H_{eff}^{(0)}={\vec{p}^2\over 2m}+e\phi
\end{equation}
and there is {\it no} interaction with the field.  In higher orders, 
there {\it can} exist such atom-field interactions and, in order to 
construct the
effective interaction, we demand that the
Hamiltonian satisfy fundamental symmetry requirements.  In
particular $H_{eff}$ must be gauge invariant, must be a scalar under
rotations, and must be even under both parity and time reversal 
transformations.  Also,
since we are dealing with Compton scattering, $H_{eff}$ must be
quadratic in the vector potential.
Actually, from the requirement of gauge invariance it is
clear that the effective interaction should involve only the electric and magnetic
fields
\begin{equation}
\vec{E}=-\vec{\nabla}\phi-{\partial\over \partial t}\vec{A}, 
\qquad \vec{B}=\vec{\nabla}\times\vec{A}\label{eq:ii}
\end{equation}
since these are invariant under a gauge transformation
\begin{equation}
\phi\rightarrow\phi+{\partial\over \partial t}\Lambda,\qquad \vec{A}
\rightarrow\vec{A}-\vec{\nabla}\Lambda
\end{equation}
while the vector and/or scalar potentials themselves are not.  The lowest order
interaction then can involve only the rotational invariants 
$\vec{E}^2,\vec{B}^2$
and $\vec{E}\cdot\vec{B}$.  However, under spatial
inversion---$\vec{r}\rightarrow -\vec{r}$---electric and magnetic
fields behave oppositely---$\vec{E}\rightarrow -\vec{E}$ while
$\vec{B}\rightarrow\vec{B}$---so that parity invariance rules out any
dependence on $\vec{E}\cdot\vec{B}$.  Likewise under time
reversal---$t\rightarrow -t$ we have $\vec{E}\rightarrow \vec{E}$ but
$\vec{B}\rightarrow -\vec{B}$ so such a term is also ruled out by time
reversal invariance.  The simplest such effective Hamiltonian must
then have the form
\begin{equation}
H_{eff}^{(1)}=-{1\over 2}c_E\vec{E}^2
-{1\over 2}c_B\vec{B}^2\label{eq:ll}
\end{equation}
(Forms involving time or spatial derivatives are much smaller.)
We know from electrodynamics that 
${1\over 2}(\vec{E}^2+\vec{B}^2)$
represents the field energy per unit volume so, by dimensional
arguments, in order to represent an
energy in Eq. \ref{eq:ll}, $c_E,c_B$ must have dimensions of volume.
Also, since the photon has such a
long wavelength, there is no penetration of the atom, so  only classical scattering
is allowed.  The relevant scale must then be atomic size so that we can write
\begin{equation}
c_E=k_Ea_0^3,\qquad c_B=k_Ba_0^3
\end{equation}
where we expect $k_E,k_B\sim {\cal O}(1)$.  Finally, since for photons
with polarization $\hat{\epsilon}$ and four-momentum $q_\mu$ we
identify $\vec{A}(x)=\hat{\epsilon}\exp(-iq\cdot x)$
then from Eq. \ref{eq:ii}, $|\vec{E}|\sim \omega$, 
$|\vec{B}|\sim |\vec{k}|=\omega$ and 
\begin{equation}
{d\sigma\over d\Omega}\propto|<f|H_{eff}|i>|^2\sim\omega^4 a_0^6
\end{equation}
as found in the previous section via detailed calculation.

\subsection{QCD}
 
Now let's return to the relevance of effective field theory to QCD.  
In many ways 
QCD represents the antithesis of QED, which can reliably be evaluated via
ordinary perturbative methods to any given order in the coupling constant. 
As mentioned previously, one of the problems with applications of 
QCD at low energy is that it is written in terms
of quark-gluon degrees of freedom rather than in terms of hadrons, with which 
experiments are done.  In addition, because of asymptotic freedom,
the effective quark-gluon coupling constant is {\it large} at low energies,
meaning that traditional perturbative methods for solving the theory will 
not work.  
Finally, since gluons interact with each other via these large coupling, 
the theory is hopelessly nonlinear and one does not then have an exact 
solution from which an effective interaction can be derived in the low 
energy limit.  The resolution of these problems is to construct an 
{\it effective} theory in terms of hadronic degrees of 
freedom which {\it in the low energy regime} matches onto the predictions 
of QCD, just as the effective interaction for Rayleigh scattering gave a 
fully satisfactory description of photon-atom scattering in the limit of 
small photon energy.  In the optical scattering case we were able to 
deduce the form of the effective 
Lagrangian by requiring that it satisfy certain general principles.  
The same will be true for QCD---we demand that the low 
energy effective Lagrangian, written in terms of meson 
($\phi$) and baryon ($\psi$) degrees of freedom---${\cal L}(\phi,\psi)$---have 
the same {\it symmetries} as does ${\cal L}_{QCD}(q,\bar{q},A_\mu^i)$, in
particular (broken) chiral symmetry.  
    
\section{Chiral Symmetry and Effective Lagrangians}

The importance of symmetry in physics arises from Noether's theorem
which states that for every symmetry of the Hamiltonian there exists a 
corresponding conservation law and, associated with each such invariance, 
there is in general a related current 
$j_\mu$ which is conserved---{\it i.e.} $\partial^\mu j_\mu=0$.  This 
guarantees that the associated charge will be time-independent, since
\begin{equation}
{dQ\over dt}=\int d^3x{\partial j_0\over \partial t}=
-\int d^3x \vec{\nabla}\cdot\vec{j}
=\int_{\rm surf}d\vec{S}\cdot\vec{j}=0 
\end{equation}
where we have used Gauss' theorem and the assumption that any fields are local.

Since in quantum mechanics the time development of an operator 
$\hat{Q}$ is given by
\begin{equation}
{d\hat{Q}\over dt}=i[\hat{H},\hat{Q}]
\end{equation}
we see that such a conserved charge must commute with the Hamiltonian.  
Now it is usually the case that the symmetry is realized in a Wigner-Weyl 
fashion wherein the vacuum (or lowest energy state) of the theory, 
which satisfies $\hat{H}|0>=E_0|0>$,
is unique and has the property $\hat{Q}|0>=|0>$ 
since $\hat{H}(\hat{Q}|0>)=\hat{Q}(\hat{H}|0>)=E_0\hat{Q}|0>$.  
However, there exist in general a set of degenerate excited states 
which mix with each other under application of the symmetry charge.  
A familiar example of a Wigner-Weyl symmetry is isospin or SU(2) 
invariance.  Because this is a 
(approximate) symmetry of the Hamiltonian, particles appear in multiplets 
such as $p,n$ or $\pi^+\pi^0\pi^-$ having identical spin-parity and 
(almost) the same mass and transform into one another under
application of the isospin charges $\vec{I}$.  However, this is not the only 
situation which occurs in nature.  It is also possible (and in fact often 
the case) that the ground state of a system does {\it not} possess the same 
symmetry as does the Hamiltonian, in which case we say that the symmetry is 
realized in a Nambu-Goldstone fashion and is {\it spontaneously broken}.  
This phenomenon can even arise in classical 
mechanics.  A familiar example involves a flexible rod under compression.  
Once the compressive force exceeds a critical value, the rod will flex, 
clearly breaking the axial symmetry which characterizes it.  
(Interestingly this
particular problem was first studied by Euler in the context of beam 
bending!)  Its connection with QCD is 
studied in the next section.
    
\subsection{Spontaneous Symmetry Breaking}
 
\bigskip
{\bf Chiral Symmetry}\par\bigskip
In order to understand the relevance of spontaneous symmetry 
breaking to QCD,
we introduce the idea of "chirality," defined by the operators
\begin{equation} \Gamma_{L,R} = {1\over 2}(1\pm\gamma_5)={1\over 2}
\left( \begin{array}{c c }
1 & \mp 1 \\
\mp 1 & 1
\end{array}\right)
\end{equation}
which project left- and right-handed components of the Dirac wavefunction
via\footnote{Note that in the limit of vanishing mass, chirality becomes
identical with helicity.}
\begin{equation} \psi_L = \Gamma_L \psi \qquad \psi_R=\Gamma_R
\psi \quad\mbox{with}\quad \psi=\psi_L+\psi_R \end{equation}
In terms of chirality states the quark component of the QCD Lagrangian
can be written as
\begin{equation} \bar{q}(i\not\! \! D-m)q=\bar{q}_Li\not \! \! D q_L +
\bar{q}_Ri\not\!\! D q_R -\bar{q}_L m q_R-\bar{q}_R m q_L 
\end{equation}
We observe then that in the limit as  
$m\rightarrow0$ 
\begin{equation} {\cal L}_{\rm QCD}\stackrel{m\rightarrow 0}
{\longrightarrow}\bar{q}_L i \not\!\! D q_L +
\bar{q}_R i \not\!\! D q_R -{1\over 2}{\rm tr}G_{\mu\nu}G^{\mu\nu}
\end{equation}
{\it i.e.}, the QCD Lagrangian is invariant under {\it independent} global
left- and right-handed rotations of the light (u,d,s) quarks
\begin{equation}
q_L  \rightarrow \exp (i \sum_j \lambda_j\alpha_j)
q_L,\qquad
q_R  \rightarrow \exp (i\sum_j \lambda_j \beta_j)
q_R
\end{equation}
This symmetry is called
$SU(3)_L \bigotimes SU(3)_R$ or chiral $SU(3)\times SU(3)$.  Continuing
to neglect the light quark masses,
we see that in a chiral symmetric world one would expect to have have 
sixteen---eight
left-handed and eight right-handed---conserved Noether currents
\begin{equation} \bar{q}_L\gamma_{\mu} {1\over 2} \lambda_i q_L \, ,
\qquad \bar{q}_R\gamma_{\mu}{1\over 2}\lambda_i
q_R \end{equation}
Equivalently, by taking the sum and difference, we would have eight 
conserved vector and
eight conserved axial-vector currents
\begin{equation}
V^i_{\mu}=\bar{q}\gamma_{\mu} {1\over 2}
\lambda_i q,\qquad
A^i_{\mu}=\bar{q}\gamma_{\mu}\gamma_5
 {1\over 2} \lambda_i q
\end{equation}
In the vector case, this is just a simple generalization of 
isospin (SU(2)) invariance 
to the case of SU(3).  There are 
{\it eight} ($3^2-1$) time-independent charges
\begin{equation} F_i=\int d^3 x V^i_0(\vec{x},t) \end{equation}
and there exist various supermultiplets of particles having identical 
spin-parity and
(approximately) the same mass in configurations---singlet, 
octet, decuplet, 
{\it etc.} demanded by SU(3)invariance.

If chiral symmetry were realized in the conventional (Wigner-Weyl) 
fashion one would
expect then there should also exist corresponding nearly degenerate
but {\it opposite} parity single particle states generated
by the action of the time-independent axial charges
$F^{5}_i= \int d^3 xA^i_0(\vec{x},t)$.  Indeed since
\begin{eqnarray}
H|P\rangle &= & E_P|P\rangle \nonumber \\
H(Q_5|P\rangle)&=&Q_5(H|P\rangle)
=  E_P(Q_5|P\rangle)
\end{eqnarray}
we see that $Q_5|P\rangle$ must also be an eigenstate of the Hamiltonian
with the same eigenvalue as $|P>$, which would seem to require the existence of
parity doublets.  However, experimentally this does not appear to be the
case.  Indeed although the $J^p={1\over 2}^+$ nucleon has a mass of about 
1 GeV, the nearest ${1\over 2}^-$ resonce lies nearly 600 MeV higher in 
energy.  Likewise in
the case of the $0^-$ pion which has a mass of about 140 MeV, the nearest 
corresponding $0^+$ state (if it exists at all) is considerably higher 
in energy. 

\subsection{Goldstone's Theorem}

One can resolve this apparent paradox by postulating that parity-doubling is
avoided because the axial symmetry is realized in a Nambu-Goldstone
fashion and is
{\it spontaneously broken.}  Then according to a theorem due to Goldstone,
when a (continuous) symmetry is broken in this fashion there must also be
generated a {\it massless} boson having the quantum
numbers of the broken generator---in this case a pseudoscalar---and when the
axial charge acts on a single particle eigenstate one does not get a single
particle eigenstate of opposite parity in return.\cite{gold} Rather 
one generates one or more of these massless pseudoscalar particles
\begin{equation} 
Q_5|P\rangle \sim |Pa \rangle +\cdots 
\end{equation}
and the interactions of such "Goldstone bosons" with each other and with other 
particles is
required to vanish as the four-momentum goes to zero.  

Now back to QCD:  According to Goldstone's argument, one would expect 
there to exist eight {\it massless} pseudoscalar states---one for each 
spontaneously broken SU(3) axial generator, which would be the Goldstone bosons of QCD.
Examination of the particle data tables reveals, however, that no such massless $0^-$ particles
exist.  There do exist eight $0^-$ particles---$\pi^\pm,\pi^0,K^\pm,K^0,\bar{K}^0\eta$
which are much lighter than their hadronic siblings.  However, these states are
certainly not massless and this causes us to ask what has gone wrong with what 
appears to be rigorous reasoning.  The answer
is found in the feature that our discussion thus far has neglected the piece
of the QCD Lagrangian associated with quark, which can be
written in the form
\begin{equation} {\cal L}_{\mbox{QCD}}^m=-(\bar{q}_Lm q_R+\bar{q}_R
mq_L)  \end{equation}
Since clearly this term breaks chiral symmetry---
\begin{eqnarray}
\bar{q}_Lm q_R &\rightarrow & \bar{q}_L \exp(-i
\sum_j \lambda_j\alpha_j) m \exp(i\sum_j \lambda_j
\beta_j) q_R \nonumber \\
& \neq & \bar{q}_Lm q_R
\end{eqnarray}
---we have a violation of the conditions under which 
Goldstone's theorem applies.  
The associated pseudoscalar bosons are {\it not} required to be massless
\begin{equation} m^2_G \neq 0 \end{equation}
but since their mass arises only from the breaking of the symmetry the
various "would-be" Goldstone masses are expected to be proportional to
the symmetry breaking parameters
\[ m^2_G \propto m_u,m_d,m_s \]
To the extent that such quark masses are small 
the eight pseudoscalar masses are not required to be massless, merely 
much lighter than other hadronic masses in the spectrum, as found in nature.

\subsection{Effective Chiral Lagrangian}

The existence of a set of particles---the pseudoscalar mesons---which are 
notably less massive than other hadrons suggests the possibility of generating
an effective field theory which correctly incorporates the chiral 
symmetry of the
underlying QCD Lagrangian in describing the low energy interactions of these
would-be Goldstone bosons.  This can
be formulated in a variety of ways, but the most transparent is done by 
including the
Goldstone modes in terms of the argument of an exponential.  Considering first
SU(2), we define 
$U=\exp(i\vec{\tau}\cdot\phi/v)$ where $\vec{\phi}$ is the pion field and
$v$ is a constant.  Then under
the chiral transformations
\begin{eqnarray}
\psi_L & \rightarrow & L \psi_L \nonumber \\
\psi_R & \rightarrow & R \psi_R
\end{eqnarray}
we have
\begin{equation} U \rightarrow L U R ^{\dagger}, \end{equation}
and a form such as
\begin{equation} \mbox{Tr} \partial^{\mu} U \partial_{\mu} U^{\dagger}
\rightarrow \mbox{Tr} L \partial^{\mu} U R^{\dagger} R
\partial_{\mu} U^{\dagger} L^{\dagger} = \mbox{Tr} \partial^{\mu}
U \partial_{\mu} U^{\dagger}\,   
\end{equation}
is invariant under chiral rotations and can be used as part of the
effective Lagrangian.  However, this form is
{\it not} one which describes realistic
Goldstone interactions since, consistent with Goldstone's theorem,
such an invariant Lagrangian must also have zero pion mass, in
contradiction to experiment.  We {\it must} then add 
a term which accounts for quark masses in order to generate 
chiral symmetry breaking and thereby non-zero pion
mass.  

In this way, we infer that the {\it lowest order} effective chiral Lagrangian 
can be written as
\begin{equation}
 {\cal L}_2={v^2 \over 4} \mbox{Tr} (\partial_{\mu}U \partial^{\mu}
 U^{\dagger})+{m^2_{\pi}\over 4} v^2 \mbox{Tr} (U+U^{\dagger})\,  .\label{eq:abc}
\end{equation}
where the subscript 2 indicates that we are working at two-derivative order
or one power of chiral symmetry breaking---{\it i.e.} $m_\pi^2$.
This Lagrangian is also {\it unique}, since if we expand to lowest order in
$\vec\phi$
\begin{equation}
\mbox{Tr}\partial_{\mu} U \partial^{\mu} U^{\dagger} =
\mbox{Tr} {i\over v} \vec{\tau}\cdot\partial_{\mu}\vec{\phi} \times
{-i\over v}\vec{\tau}\cdot\partial^{\mu}\vec{\phi}= {2\over v^2}
\partial_{\mu}\vec{\phi}\cdot \partial^{\mu}\vec{\phi}\,  ,
\end{equation}
we must reproduce the free pion Lagrangian
\begin{equation}
 {\cal L}_2 ={1\over 2} \partial_{\mu}
\vec{\phi}\cdot \partial^{\mu} \vec{\phi} -{1\over 2} m^2_{\pi}
\vec{\phi}\cdot \vec{\phi} +{\cal O} (\phi^4) \,  .
\end{equation}

At the SU(3) level, including a generalized chiral symmetry breaking term,
there is even predictive power---one has
\begin{equation}
 {v^2\over 4} \mbox{Tr} \partial_{\mu} U \partial^{\mu} U^{\dagger}
=   {1\over 2} \sum_{j=1}^8 \partial_{\mu}
\phi_j\partial^{\mu}\phi_j +\cdots \nonumber\\
\end{equation}
and 
\begin{eqnarray}
&&{v^2 \over 4} \mbox{Tr} 2 B_0 m ( U+ U^{\dagger})
 =  \mbox{const.}
-{1\over 2} (m_u+ m_d)B_0 \sum_{j=1}^3 \phi^2_j \nonumber\\
&-&{1\over 4} (m_u+m_d+2m_s)B_0\sum_{j=4}^7 \phi^2_j
 -{1\over 6} (m_u+m_d +4m_s)B_0\phi^2_8  +\cdots 
\end{eqnarray}
where $B_0$ is a constant and $m$ is the quark mass matrix. We can
then identify the meson masses as
\begin{eqnarray}
 m^2_{\pi} & =&  2\hat{m} B_0
\nonumber \\
 m_K^2 &=& (\hat{m} +m_s) B_0 \nonumber \\
m_{\eta}^2 & =& {2\over 3} (\hat{m} + 2m_s) B_0  \, ,
\end{eqnarray}
where $\hat{m}={1\over 2}(m_u+m_d)$ is the mean light quark mass.
This system of three equations is {\it overdetermined}, and we find by simple
algebra
\begin{equation}
3m_{\eta}^2 +m_{\pi}^2 - 4m_K^2 =0 \, \, .
\end{equation}
which is the Gell-Mann-Okubo mass relation and is well-satisfied
experimentally.\cite{gmo}
 
Such effective interaction calculations in the meson sector have been 
developed to a high degree over the past fifteen or so years, including the
inclusion of loop contributions, in order to preserve crossing symmetry and
unitarity.  When such loop corrections are included one must augment the
effective lagrangian to include ``four-derivative'' terms with arbitrary
coefficients which must be fixed from experiment.  This program, called
chiral perturbation theory, has been enormously successful in describing
low energy interactions in the meson sector\cite{cpt}.  However, 
in order to discuss
hyperons, we must extend it to consider baryons.

\section{Baryon Chiral Perturbation Theory}

Our discussion of chiral techniques given above was limited to the study of
the interactions of the pseudoscalar mesons (would-be Goldstone bosons) 
with leptons and with each
other.  In the real world, of course, interactions with baryons also
take place and it becomes an important problem to develop a useful predictive
scheme based on chiral invariance for such processes.  Again much work
has been done in this regard\cite{gss}, but there remain important 
problems\cite{bkm}.  Writing
down the lowest order chiral Lagrangian at the SU(2) level is
straightforward---
\begin{equation}
{\cal L}_{\pi N}=\bar{N}(i\not\!\!{D}-m_N+{g_A\over 2}\rlap /{u}\gamma_5)N
\end{equation}
where $g_A$ is the usual nucleon axial coupling in the chiral limit, the
covariant derivative $D_\mu=\partial_\mu+\Gamma_\mu$ is given by
\begin{equation}
\Gamma_\mu={1\over 2}[u^\dagger,\partial_\mu u]-{i\over 2}u^\dagger
(V_\mu+A_\mu)u-{i\over 2}u(V_\mu-A_\mu)u^\dagger ,
\end{equation}
and $u_\mu$ represents the axial structure
\begin{equation}
u_\mu=iu^\dagger\nabla_\mu Uu^\dagger
\end{equation}
The quantities $V_\mu,\,A_\mu$ represent external (non-dynamical) vector,
axial-vector fields.
Expanding to lowest order we find
\begin{eqnarray}
{\cal L}_{\pi N}&=&\bar{N}(i\rlap /{\partial}-m_N)N+g_A
\bar{N}\gamma^\mu\gamma_5{1\over 2}\vec{\tau}N\cdot({i\over F_\pi}\partial_\mu\vec{\pi}
+2\vec{A}_\mu)\nonumber\\
&-&{1\over 4F_\pi^2}\bar{N}\gamma^\mu\vec{\tau}N\cdot\vec{\pi}\times
\partial_\mu\vec{\pi}+\ldots
\end{eqnarray}
which yields the Goldberger-Treiman relation,
connecting strong and axial couplings of the nucleon system\cite{gt}
\begin{equation}
F_\pi g_{\pi NN}=m_N g_A
\end{equation}
Using the present best values for these quantities, we find
\begin{equation}
92.4 \mbox{MeV}\times 13.05 =1206 \mbox{MeV}\quad\mbox{vs.}\quad 1189 \mbox{MeV}
= 939\mbox{MeV}\times 1.267
\end{equation}
and the agreement to better than two percent strongly confirms the validity
of chiral symmetry in the nucleon sector.  Actually the Goldberger--Treiman relation is only strictly true at the off-shell point $g_{\pi NN}(q^2=0)$ rather
than at the physical value of the coupling $g_{\pi NN}(q^2=m_\pi^2)$
and one {\it expects}
$\sim 2\%$ discrepancy to exist.  An interesting "wrinkle" in this regard
is the use of the so-called Dashen-Weinstein relation which uses simple SU(3)
symmetry breaking to predict this discrepancy in terms of corresponding numbers in the
strangeness changing sector.\cite{dw}

\subsection{Heavy Baryon Methods}

Extension to SU(3) gives additional successful predictions---the baryon
Gell-Mann-Okubo formula as well as the generalized Goldberger-Treiman
relation.  However, difficulties arise when one attempts to include
higher order corrections to this formalism.  The difference from the
Goldstone case is that there now exist {\it three} dimensionful
parameters---$m_N$, $m_\pi$ and $F_\pi$---in the problem rather than
just $m_\pi$ and $F_\pi$.  Thus loop effects can be of order 
$(m_N/4\pi F_\pi)^2
\sim 1$ and we no longer have a reliable perturbative scheme.  A
consistent power counting mechanism {\it can} be constructed provided that we
eliminate the nucleon mass from the leading order Lagrangian.  This is done by
considering the nucleon to be very heavy.  Then we can write its
four-momentum as\cite{jm}
\begin{equation}
p_\mu=Mv_\mu+k_\mu
\end{equation}
where $v_\mu$ is the four-velocity and satisfies $v^2=1$, while $k_\mu$
is a small off-shell momentum, with $v\cdot k<< M$.  One can construct
eigenstates of the projection operators $P_\pm = {1\over 2}(1\pm
\rlap /{v})$, which in the rest frame select upper, lower
components of the Dirac wavefunction, so that\cite{bkkm}
\begin{equation}
\psi=e^{-iMv\cdot x}(H_v+h_v)
\end{equation}
where
\begin{equation}
H_v=P_+\psi,\qquad h_v=P_-\psi
\end{equation}
The $\pi N$ Lagrangian can then be written in terms of $H,h$ as
\begin{equation}
{\cal L}_{\pi N}=\bar{H}_v{\cal A}H_v+\bar{h}_v{\cal B}H_v+
\bar{H}_v\gamma_0{\cal B}^\dagger\gamma_0h_v-\bar{h}_v{\cal C}h_v\label{eq:hh}
\end{equation}
where the operators ${\cal A}, {\cal B},{\cal C}$ have the low energy
expansions
\begin{eqnarray}
{\cal A}&=&iv\cdot D+g_A u\cdot S +\ldots\nonumber\\
{\cal B}&=&i\not\!\!{D}^\perp-{1\over 2}g_A v\cdot u\gamma_5+\ldots\nonumber\\
{\cal C}&=&2M+iv\cdot D+g_A u\cdot S+\ldots
\end{eqnarray}
Here $D_\mu^\perp=(g_{\mu\nu}-v_\mu v_\nu)D^\nu$ is the transverse component
of the covariant derivative and $S_\mu={i\over 2}\gamma_5
\sigma_{\mu\nu}v^\nu$ is the Pauli-Lubanski spin vector and satisfies
\begin{equation}
S\cdot v=0,\quad S^2=-{3\over 4},\quad\{S_\mu,S_\nu\}={1\over 2}(v_\mu v_\nu-
g_{\mu\nu}),\quad [S_\mu,S_\nu]=i\epsilon_{\mu\nu\alpha\beta}v^\alpha S^\beta
\end{equation}
We observe that the two components H,h are coupled Eq. \ref{eq:hh}.
However, the system may be diagonalized by use of 
the field transformation
\begin{equation}
h'=h-{\cal C}^{-1}{\cal B}H
\end{equation}
in which case the Lagrangian becomes
\begin{equation}
{\cal L}_{\pi N}=\bar{H}_v({\cal A}+(\gamma_0{\cal B}^\dagger
\gamma_0){\cal C}^{-1}{\cal B})H_v-\bar{h}'_v{\cal C}h'_v
\end{equation}
The piece of the Lagrangian involving $H$ contains the mass only in 
the operator
${\cal C}^{-1}$ and is the effective Lagrangian that we seek.  The remaining
piece involving $h'_v$ can be thrown away, as it does not couple to the
$H_v$ physics.  (In path integral language we simply integrate out this
component yielding an uninteresting overall constant.)
Of course, when loops are included a set of
counterterms will be required and these are given at leading (two-derivative)
order by
\begin{eqnarray}
{\cal A}^{(2)}&=&{M\over F_\pi^2}(c_1\mbox{Tr}\chi_+
+c_2(v\cdot u)^2+c_3u\cdot u+c_4[S^\mu,s^\nu]u_\mu u_\nu\nonumber\\
&+&c_5(\chi_+-\mbox{Tr}\chi_+)
-{i\over
4M}[S^\mu,S^\nu]((1+c_6)F^+_{\mu\nu}+c_7\mbox{Tr}f^+_{\mu\nu}))\nonumber\\
{\cal B}^{(2)}&=&{M\over F_\pi^2}((-{c_2\over 4}i[u^\mu,u^\nu]+c_6f_+^{\mu\nu}
+c_7Trf_+^{\mu\nu})\sigma_{\mu\nu}\nonumber\\
&-&{c_4\over 2}v_\mu\gamma_\nu Tru^\mu u^\nu)\nonumber\\
{\cal C}^{(2)}&=&-{M\over F_\pi^2}(c_1Tr\chi_++(-{c_2\over 4}i[u^\mu,u^\nu]
+c_6f_+^{\mu\nu}+c_7trF_+^{\mu\nu})\sigma_{\mu\nu}\nonumber\\
&-&{c_3\over 4}Tr u^\mu u_\nu
-({c_4\over 2}+Mc_5)v_\mu v_\nu Tru^\mu u^\nu)\label{eq:de}
\end{eqnarray}
where $f^{\pm}_{\mu\nu}=u^\dagger F^R_{\mu\nu}u\pm uF^L_{\mu\nu}u^\dagger$ and 
$\chi_+=umu+u^\dagger mu^\dagger$.
Expanding ${\cal C}^{-1}$ and the other terms in terms of
a power series in $1/M$ leads to an
effective heavy nucleon Lagrangian of the form (to ${\cal O}(q^3)$)
\begin{eqnarray}
{\cal L}_{\pi N}&=&\bar{H}_v\{{\cal A}^{(1)}+{\cal A}^{(2)}+{\cal A}^{(3)}
+(\gamma_0{\cal B}^{(1)\dagger}\gamma_0){1\over 2M}{\cal B}^{(1)}\nonumber\\
&+&{(\gamma_0{\cal B}^{(1)\dagger}\gamma_0){\cal B}^{(2)}+(\gamma_0{\cal B}
^{(2)\dagger}\gamma_0){\cal B}^{(1)}\over 2M}\nonumber\\
&-&(\gamma_0{\cal B}^{(1)\dagger}\gamma_0){i(v\cdot D)+g_A(u\cdot S)\over
(2M)^2}{\cal B}^{(1)}\}H_v+{\cal O}(q^4)\label{eq:def}
\end{eqnarray}
A set of Feynman rules can now be written down and a consistent power counting
scheme developed, as shown by Meissner and his collaborators.\cite{bkm}

\subsection{Applications}
As an example of the use of this formalism, called heavy baryon chiral perturbation
theory (HB$\chi$pt) consider the nucleon-photon interaction.  To lowest (one
derivative) order we have from ${\cal A}^{(1)}$
\begin{equation}
{\cal L}_{\gamma NN}^{(1)}=ie\bar{N}{1\over2}(1+\tau_3)\epsilon\cdot vN
\end{equation}
while at two-derivative level we find
\begin{eqnarray}
 {\cal L}_{\gamma NN}^{(2)}=\bar{N}\left\{{e\over
4M}(1+\tau_3)\epsilon\cdot(p_1+p_2)
+{ie\over 2M}[S\cdot \epsilon,S\cdot k](1+\kappa_S+\tau_3(1+\kappa_V)\right\}N
\nonumber\\
\quad
\end{eqnarray}
where we have made the identifications
$c_6=\kappa_V,\,\,c_7={1\over 2}(\kappa_S-\kappa_V)$.  We can now reproduce
the low energy theorems for Compton scattering.  Consider the case of the
proton. At the two derivative level,
we have the tree level prediction 
\begin{equation}
(\gamma_0{\cal B}^{(1)\dagger}\gamma_0){1\over 2M}
{\cal B}^{(1)}|_{\gamma pp}={e^2\over 2M}\vec{A}_\perp^2
\end{equation}
which yields the familiar Thomson amplitude
\begin{equation}
\mbox{Amp}_{\gamma pp}=-{e^2\over M}\hat{\epsilon}'\cdot\hat{\epsilon}
\end{equation}

Again an entire literature on this subject exists, and we shall have
to be content with only a brief introduction.  It is now time to return to the 
hyperon sector to see how such chiral ideas can been applied.

\section{Hyperon Properties}
The quantum numbers of the ground state hyperons are easily described
by noting that they, together with the nucleons, constitute an SU(3) octet.
Alternatively, these values follow directly from the underlying
three-quark structure which, together with the masses and 
magnetic moments, is listed in Table 1.
\begin{table}
\begin{center}
\begin{tabular}{cccc}
Hyperon&Quark Structure& Mass & Magnetic Moment\\
$\Lambda$&uds&1115.684(6)&-0.613(4)\\
$\Sigma^+$&uus&1189.37(7)&2.458(10)\\
$\Sigma^0$&uds&1192.55(8)& \\
$\Sigma^-$&dds&1197.45(4)&-1.160(25)\\
$\Xi^0$&uss&1314.9(6)&-1.250(14)\\
$\Xi^-$&dss&1321.32(13)&-0.6507(25)\\
$\Omega^-$&sss&1672.5(7)&-2.02(5)
\end{tabular}
\caption{Basic Hyperon Properties.  Here masses are given in MeV and 
magnetic moment in nucleon magnetons.}
\end{center}
\end{table}
One can achieve a basic understanding of these masses either by use
of a constituent quark picture or via symmetry methods.  In the former one
employs the simple quark mass matrix together with an effective hyperfine
interaction arising from gluon exchange
\begin{equation}
{\cal H}_{hyp}={1\over 2}\sum_{i>j}
C_{ij}\vec{s}_i\cdot\vec{s}_j\delta^3(\vec{r}_i-\vec{r}_j)
\end{equation}
and finds, {\it e.g.}, in a simple bag model,
\begin{eqnarray}
M_N&=&3m_n\int d^3r(u^2-\ell^2)-{3\over 8}C_{nn}\nonumber\\
M_\Lambda&=&(2m_n+m_s)\int d^3r(u^2-\ell^2)-{3\over 8}C_{nn}\nonumber\\
M_\Sigma&=&(2m_n+m_s)\int d^3r(u^2-\ell^2)+{1\over 8}C_{nn}-{1\over
2}C_{ns}\nonumber\\
M_\Xi&=&(2m_s+m_n)\int d^3r(u^2-\ell^2)-{1\over 2}C_{ns}
+{1\over 8}C_{ss}
\end{eqnarray}
where the subscripts $n,s$ refer to strange,nonstrange respectively
and $u,\ell$ refer to the upper, lower components of the bag
wavefunction.

Alternatively, one can use a symmetry-based approach.  
An effective SU(3) heavy baryon strong interaction chiral 
lagrangian for the hyperon 
system can be written in the form
\begin{eqnarray}
{\cal L}&=&{\rm tr}\bar{B}iv\cdot DB+d_A{\rm tr}\bar{B}S^\mu\{u_\mu,B\}
\nonumber\\
&+&f_A{\rm tr}\bar{B}S^\mu[u_\mu,B]+d_m{\rm tr}\bar{B}\{\chi_+,B\}\nonumber\\
&+&f_m{\rm tr}\bar{B}[\chi_+,B]+b_0{\rm tr}\bar{B}B{\rm tr}\chi_++\ldots
\end{eqnarray}
where $\chi_+$ represents the symmetry breaking due
to the non-zero quark mass $m$ and has been defined below Eq. \ref{eq:de}. 
At tree level one finds
\begin{eqnarray}
M_N&=&\bar{M}_0+4(m_K^2-m_\pi^2)f_m-4m_K^2d_m\nonumber\\
M_\Lambda&=&\bar{M}_0-{4\over 3}(4m_K^2-m_\pi^2)d_m\nonumber\\
M_\Sigma&=&\bar{M}_0+-4m_\pi^2d_m\nonumber\\
M_\Xi&=&\bar{M}_0-4(m_K^2-m_\pi^2)f_m-4m_K^2d_m
\end{eqnarray}
where $\bar{M}_0=M_0-2(2m_K^2+m_\pi^2)b_0.$

In either case we find a sum rule---the Gell-Mann-Okubo 
relation\footnote{Note that
in the quark model case we assume the first order symmetry
breaking relation $C_{nn}-C_{ns}=C_{ns}-C_{ss}$.}
\begin{equation}
M_\Sigma-M_N={1\over 2}(M_\Xi-M_N)+{3\over 4}(M_\Sigma-M_\Lambda)
\end{equation}
which is well-satisfied experimentally---254 MeV vs. 248 MeV\cite{gmo}.

These results have been known for well over three decades.  However,
in contemporary treatments, as described above, 
it has become traditional to include higher 
order chiral contributions, and this is where one runs into 
difficulties---at one loop in HB$\chi$pt the nonanalytic corrections are
exactly calculable and very sizable
\begin{eqnarray}
\delta M_N&=&-0.31\,\,{\rm GeV},\quad \delta M_\Sigma=-0.62\,\,
{\rm GeV}\nonumber\\
\delta M_\Lambda&=&-0.66\,\, {\rm GeV},\quad \delta M_\Xi=-1.02\,\,
{\rm GeV}\label{eq;abc}
\end{eqnarray}
so that, {\it e.g.} the $\Xi$ mass receives a 100\% correction!  Of course,
this is not a fatal flaw since such large nonanalytic corrections can be 
compensated by inclusion of
appropriate higher order counterterms.  Indeed the 
calculation has been extended to ${\cal O}(p^4)$ by Borasoy and Meissner, who
obtained\cite{bmeis}
\begin{eqnarray}
M_N&=&\bar{M}_0(1+0.34-0.35+0.24)\nonumber\\
M_\Sigma&=&\bar{M}_0(1+0.81-0.70+0.44)\nonumber\\
M_\Lambda&=&\bar{M}_0(1+0.69-0.77+0.54)\nonumber\\
M_\Xi&=&\bar{M}_0(1+1.10-1.16+0.78)\label{eq;bcd}
\end{eqnarray}
where the nonleading terms refer to the contribution from ${\cal O}(p^2)$
counterterms, nonanalytic pieces of ${\cal O}(p^3)$, and ${\cal O}(p^4)$
counterterms respectively.  However, although the masses can be fit in
such a picture, the structure of the series generates  
concern with respect to convergence.  Also in this approach the 
Gell-Mann-Okubo discrepancy is predicted to be about 30 MeV---a factor of 
five larger than the 6 MeV found experimentally!.  

One way out of this conundrum has been suggested by Donoghue and Holstein
\cite{dhol}, who
point out that the origin of these large higher order corrections is the
form of the ${\cal O}(p^3)$ nonanalytic terms, which arise from the 
HB$\chi$pt integral
\begin{equation}
\int {d^4k\over (2\pi)^4}{k_ik_j\over (k_0-i\epsilon)(k^2-m^2+i\epsilon)}=
-i\delta_{ij}{I(m)\over 24\pi}
\end{equation}
If one uses dimensional regularization to evaluate this integral, the result is
\begin{equation}
I(m)_{\rm dim-reg}=m^3\label{eq:int}
\end{equation}
which leads to the large nonanalytic corrections in Eq. \ref{eq;abc},
which must be cancelled by correspondingly large 
next order counterterms.  The result is an expansion of the form ({\it cf.}
Eq. \ref{eq;bcd})
\begin{equation}
Amp=A_0(1-1+1-1+\ldots)
\end{equation}
However, {\it physics} tells us that this approach may be misleading.  The 
point is that in the real world baryons have a finite size---$\sim$1 fm---and,
while effects involving scales larger than this are no doubt correctly
described by the effective chiral lagrangian, processes associated with
smaller scales will be significantly modified by baryon structure.  On the
other hand the integration in Eq. \ref{eq:int} treats both long and
short distance effects with equal weight, and this is the origin of the
convergence problem.  A possible solution can be found by introducing 
some sort of
weighting function which is unity at long distance but which is damped 
at short distance scales, in order to simulate baryon structure.  
For analytic purposes it is simplest to utilize
a simple dipole regulator
\begin{equation}
\left({\Lambda^2\over \Lambda^2-m^2}\right)^2
\end{equation}
in which case the function $I(m)$ takes the form
\begin{equation}
I_\Lambda(m)=\Lambda^4{2m+\Lambda\over 2(m+\Lambda)^2}
\end{equation}

\begin{figure}
\begin{center}
\epsfig{file=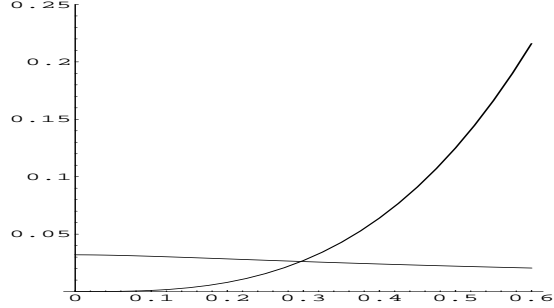,height=4cm,width=10cm}
\caption{Shown is the function $I_\Lambda(m)$ (monotonically decreasing
curve) vs. its dimensionally regularized counterpart (monotonically
increasing curve).  Obviously the influence of heavy exchanged mesons is
strongly suppressed in the long distance regularized version.}
\end{center}
\end{figure}

Notice that in the case that the mass is small compared to the cutoff $\Lambda$
\begin{equation}
I_\Lambda(m)\stackrel{m<<\Lambda}{\longrightarrow}{1\over 2}\Lambda^3
-{1\over 2}\Lambda m^2+m^3+\ldots
\end{equation}
In dimensional regularization terms involving the cutoff scale $\Lambda$ are
discarded leaving only the term in $m^3$.  However, using cutoff regularization
the terms in $\Lambda$ are retained, which emphasizes long distance
effects and eliminates their short-distance counterparts, as can be 
seen by examining the opposite limit
\begin{equation}
I_\Lambda(m)\stackrel{m>>\Lambda}{\longrightarrow}{\Lambda^4\over m}
-{3\over 2}{\Lambda^5\over m^2}+\ldots
\end{equation}
We observe that the function $I_\Lambda(m)$ then depends upon the pseudoscalar
mass via inverse powers.  Consequently, the pion will be strongly emphasized 
over its heavier eta or kaon counterparts, which is what is expected 
physically.
This can be seen clearly in Figure 2, 
which plots the integral $I_\Lambda(m)$
together with its dimensional regularized form.

The fit to the hyperon masses which results from such a program is shown in 
Table 2.  Obviously the convergence of the chiral expansion is now under 
control and the results are only weakly dependent upon the cutoff parameter.
That there exists no strong cutoff dependence results from the feature that the
divergent terms in $\Lambda^3$ and $\Lambda$ can be absorbed into 
renormalized values of effective parameters in the theory via
\begin{eqnarray}
M_0^r&=&\bar{M}_0-(5d_A^2+9f_A^2){\Lambda^3\over 48\pi F_\pi^2}\nonumber\\
d_m^r&=&d_m-(3f_A^2-d_A^2){\Lambda\over 128\pi F_\pi^2}\nonumber\\
f_m^r&=&f_m-5d_Af_A{\Lambda\over 192\pi F_\pi^2}\nonumber\\
b_0^r&=&b_0-(13d_A^2+9f_A^2){\Lambda\over 576\pi F_\pi^2}
\end{eqnarray}
Since such parameters are determined phenomenologically anyway, the
use of such renormalized parameters is not a problem.

\begin{table}
\begin{center}
\begin{tabular}{cccccc}
  & Dim.& $\Lambda=300$&$\Lambda=400$&$\Lambda=500$&$\Lambda=600$\\
N&-0.31&0.02&0.03&0.05&0.07\\
$\Sigma$&-0.62&0.03&0.05&0.08&0.12\\
$\Lambda$&-0.66&0.03&0.06&0.12&0.17\\
$\Xi$&-1.03&0.04&0.08&0.12&0.17
\end{tabular}
\caption{Nonanalytic contributions (in GeV) to baryon masses in 
dimensional regularization and for various values of the cutoff parameter
$\Lambda$ in MeV.}
\end{center}
\end{table}

That there exists cutoff dependence at all is due 
to the fact that we have
not included the actual short distance effects. If we had a 
reasonable model of
such pieces the result would, of course, be independent of renormalization
point.  The relative independence with respect to $\Lambda$ is associated
with the fact that such effects are small, as expected from physics 
considerations.  Of course, besides cutoff independence what we have lost
is the consistent chiral power counting which is one of the cornerstones 
of the chiral procedure---cutoff regularization mixes terms of 
different orders.  However, one can consider the price worth paying as it
provides a chirally consistent picture of the physics.

Although not strictly based on chiral arguments  it is important to note 
a parallel program which attempts to understand the hyperon masses---the 
$1/N_c$ approach being pursued at UCSD\cite{jenk} and which was 
motivated by the observation of large cancellations in parallel chiral
loops calculations involving octet baryon and decuplet intermediate
states.   
The basis of this procedure is the feature that in the large $N_c$ 
limit a spin-flavor symmetry exists for baryons---large $N_c$ baryons 
lie in irreducible 
representations of the spin-flavor group and one can 
calculate static properties in a systematic $1/N_c$ expansion.  More
precisely one can perform a simultaneous expansion in $1/N_c$ and in 
flavor breaking
and the resulting mass relations can be placed in a corresponding hierarchy,
whereby the higher the order in symmetry breaking and $1/N_c$ the better
such a relation should be satisfied.  An example is shown in Table 3,
which is taken from work by Jenkins and Lebed\cite{jl}.
\begin{table}
\begin{center}
\begin{tabular}{c|c|c|c}
Relation&$1/N_c$&Flavor&Expt.\\
\hline\\
$25(\bar{N}+3\bar{\Sigma}+\Lambda+2\bar{\Xi})-4(4\bar{\Delta}
+3\bar{\Sigma}^*+2\bar{\Xi}^*+\Omega)$&$N_c$&1& ...\\
$5(2\bar{N}+3\bar{\Sigma}+\bar{\Lambda}+2\bar{\Xi})-4(4\bar{\Delta}
+3\bar{\Sigma}^*+2\bar{\Xi}^*+\Omega)$&$1/N_c$&1&18.2\%\\
$5(6\bar{N}-3\bar{\Sigma}+\Lambda-4\bar{\Xi})-2(2\bar{\Delta}+3\bar{\Xi}^*
-\Omega)$&1&$\epsilon$&20.2\%\\
$\bar{N}-3\bar{\Sigma}+\Lambda+\bar{\Xi}$&$1/N_c$&$\epsilon$&5.9\%\\
$(-2\bar{N}-9\bar{\Sigma}+3\Lambda+8\bar{\Xi})+2(2\bar{\Delta}-\bar{\Xi}^*
-\Omega)$&$1/N_c^2$&$\epsilon$&1.1\%\\
$35(2\bar{N}-\bar{\Sigma}-3\Lambda+2\bar{\Xi})-4(4\bar{\Delta}-5\bar{\Sigma}^*
-2\bar{\Xi}^*+3\Omega)$&$1/N_c$&$\epsilon^2$&0.4\%\\
$7(2\bar{N}-\bar{\Sigma}-3\Lambda+2\bar{\Xi})-2(4\bar{\Delta}-5\bar{\Sigma}^*
-2\bar{\Xi}^*+3\Omega)$&$1/N_c^2$&$\epsilon^2$&0.2\%\\
$\bar{\Delta}-3\bar{\Sigma}^*+3\bar{\Xi}^*-\Omega$&$1/N_c^2$&$\epsilon^3$&0.1\%
\end{tabular}
\caption{Mass relations arising in simultaneous SU(3) and $1/N_c$ expansion.
Here the bar indicates the average mass of the multiplet.}
\end{center}
\end{table}
Since both $1/N_c$ and the flavor breaking parameter 
$\epsilon$ are expected to be of order $30\%$ the
results of this procedure are impressive and deserve a good deal
of attention\cite{le}. 
 
It is possible to understand how some of these relations come about by 
considering a $1/N_c$ approach within the chiral loop expansion.  In 
this approach
one treats $g_A$ as being ${\cal O}(N_c)$ and $F_\pi$ as being ${\cal O}
(\sqrt{N_c})$.  A baryon-meson vertex then grows as $\sqrt{N_c}$, which means
that individual meson-baryon scattering diagrams grow as $N_c$ and soon
violate unitarity unless there exist certain consistency conditions.
General arguments of $1/N_c$ counting produce cancellations among 
the diagrams contributing
to chiral mass corrections when both decuplet and octet states are included
as intermediate states, allowing the overall correction to be in line with
general arguments although individual diagrams are not.  Details of how this
comes about can be found in refs. \cite{dm},\cite{ej},\cite{dr}.  

\section{Magnetic Moments}

A second issue which has received a good deal of study over the years is
that of hyperon magnetic moments.  In this case a simple SU(3) analysis
represents the magnetic moments in terms of lowest order 
couplings $f_\mu,d_\mu$ via 
\begin{equation}
{\cal L}_{em}={-if_\mu\over 4M_0}{\rm tr}\bar{B}[S^\mu,S^\nu][F_{\mu\nu}^+,B]
-{id_\mu\over 4M_0}{\rm tr}\bar{B}[S^\mu,S^\nu]\{F_{\mu\nu}^+,B\}
\end{equation}
Equivalently, in a simple quark model the hyperon magnetic 
moments can be fit in terms of the corresponding quark quantities, as given
in Table 4.
\begin{table}
\begin{center}
\begin{tabular}{ccc}
particle moment& SU(3) & $\mu_s\neq\mu_d$\\
$\mu_p={4\mu_u-\mu_d\over 3}$&2.6&2.8\\
$\mu_n={4\mu_d-\mu_u\over 3}$&-1.6&-1.9\\
$\mu_\Lambda=\mu_s$&-0.8&-0.6\\
$\mu_{\Sigma^+}={4\mu_u-\mu_s\over 3}$&2.6&2.7\\
$\mu_{\Sigma\Lambda}={\mu_u-\mu_d\over \sqrt{3}}$&1.7&1.6\\
$\mu_{\Sigma^-}={4\mu_d-\mu_s\over 3}$&-0.8&-1.1\\
$\mu_{\Xi^0}={4\mu_s-\mu_u\over 3}$&-1.6&-1.4\\
$\mu_{\Xi^-}={4\mu_s-\mu_d\over 3}$&-0.8&-0.5
\end{tabular}
\caption{Theoretical values for hyperon magnetic moments in units of nucleon
magnetons obtained via a
global fit.  The second column
gives values in the lowest order chiral (SU(3)) limit wherein $\mu_d=\mu_s$
while the third column gives values in the SU(3) broken limit wherein
$\mu_s<\mu_d$ due to the heavier s-quark mass.}
\end{center}
\end{table}
In either case such a lowest order fit leads to the so-called 
Coleman-Glashow relations\cite{cg}
\begin{eqnarray}
\mu_{\Sigma^+}&=&\mu_p,\quad2\mu_\Lambda=\mu_n,\quad \mu_{\Sigma^-}
+\mu_n=-\mu_p\nonumber\\
\mu_{\Sigma^-}&=&\mu_{\Xi^-},\quad\mu_{\Xi^0}=\mu_n,\quad 
2\mu_{\Lambda\Sigma^0}=-\sqrt{3}\mu_n
\end{eqnarray}
as well as the isospin relation
\begin{equation}
\mu_{\Sigma^0}={1\over 2}(\mu_{\Sigma^+}+\mu_{\Sigma^-})
\end{equation}
Such relations also follow from a naive (degenerate) quark model picture of
hyperon structure and are in fair agreement with experiment.  This 
agreement can be considerably improved by including the u,d- and s-quark
mass difference into the analysis, however, as shown in the third column of
Table 4.

As in the case of the hyperon masses, the first calculation of chiral
corrections was that of the leading nonanalytic piece.  This was done by
Caldi and Pagels\cite{cp}, who found the form
\begin{equation}
\delta\mu_i={M_0\over 8\pi F_\pi^2}\sum_j\sigma_i^jm_j
\end{equation} 
where the coefficients $\sigma_j^i$ are given in ref. \cite{dhol} in terms
of $f_A,g_A$.  Corrections are large and in general 
destroy agreement with experiment.  It is interesting to note, though, that
three relations remain 
\begin{eqnarray}
-\mu_{\Sigma^+}=-2\mu_\Lambda-\mu_{\Sigma^-}
&\quad&\mu_{\Xi^0}+\mu_{\Xi^-}
+\mu_n=2\mu_\Lambda-\mu_p\nonumber\\
\mu_\Lambda-\sqrt{3}\mu_{\Lambda\Sigma^0}&=&\mu_{\Xi^0}+\mu_n
\end{eqnarray}
and are reasonably well satisfied by the data.  Of course, overall agreement 
can be restored 
by inclusion of counterterms and a complete
one loop evaluation has been done by Meissner and Steininger\cite{ms}, 
yielding 
results consistent with experiment, but also lacking predictive power---
\begin{eqnarray}
\mu_p&=&4.48(1-0.49+0.11)=2.79\nonumber\\
\mu_n&=&-2.47(1-0.34+0.12)=-1.91\nonumber\\
\mu_{\Sigma^+}&=&4.48(1-0.62+0.17)=2.46\nonumber\\
\mu_{\Sigma^-}&=&-2.01(1-0.31-0.11)=-1.16\nonumber\\
\mu_{\Sigma^0}&=&1.24(1-0.87+0.40)=0.65\nonumber\\
\mu_\Lambda&=&-1.24(1-0.87+0.37)=-0.61\nonumber\\
\mu_{\Xi^0}&=&-2.47(1-0.89+0.40)=-1.25\nonumber\\
\mu_{\Xi^-}&=&-2.01(1-0.64-0.03)=-0.65\nonumber\\
\mu_{\Lambda\Sigma^0}&=&2.14(1-0.53+0.19)
\end{eqnarray}
where, as before, the three contributions represent the leading ${\cal O}(p^2)$, 
nonanalytic ${\cal O}(p^3)$, and ${\cal O}(p^4)$ counterterm contributions
respectively.  As in the case of the hyperon mass analysis, the basic 
agreement which obtains at lowest
order is restored but convergence of the chiral series remains an issue.

One solution to this problem is to use long distance regularization.  In
this case there exists a linearly divergent component of the integral in
the limit of large $\Lambda$ which can be absorbed into renormalization of
the $f_\mu,d_\mu$ couplings via
\begin{eqnarray}
d_\mu^r&=&d_\mu+{\bar{M}_0\Lambda\over 4\pi F_\pi^2}d_Af_A\nonumber\\
f_\mu^r&=&f_\mu+{\bar{M}_0\Lambda\over 72\pi F_\pi^2}(5d_A^2+9f_A^2)
\end{eqnarray}
and when this is done the remaining chiral corrections become much smaller and
under control, as found in the analogous hyperon mass analysis.

The magnetic moments have also been examineed via $1/N_c$
methods\cite{jm1}\cite{dd}.  As in the case of the masses the 
leading terms in the 
expansion are ${\cal O}(N_c)$.  However, by analyzing the corrections 
via a simultaneous expansion in $m_s$ and $1/N_c$ one can identify 
relations which
are more generally valid.  These can be used in order to predict values for
decuplet magnetic moments.  For example, the isospin relations 
\begin{eqnarray}
2\mu_{\Sigma^0}&=&\mu_{\Sigma^+}+\mu_{\Sigma^-}\nonumber\\
\mu_{\Delta^{++}}-\mu_{\Delta^+}&=&\mu_{\Delta^0}-\mu_{\Delta^-}\nonumber\\
\mu_{\Delta^{++}}-\mu_{\Delta^-}&=&3(\mu_{\Delta^+}-\mu_{\Delta^0})\nonumber\\
2\mu_{\Sigma^{*0}}&=&\mu_{\Sigma^{*+}}+\mu_{\Sigma^{*-}}
\end{eqnarray}
are valid to all orders in $N_c$ and $m_s$, while the SU(3) identities
\begin{eqnarray}
0&=&\mu_{\Sigma^{++}}-\mu_{\Sigma^+}-\mu_{\Sigma^-}+\mu_{\Xi^0}-\mu_{\Xi^-}
\nonumber\\
0&=&\mu_{\Sigma^{*-}}-2\mu_{\Xi^{*-}}+\mu_{\Omega^-}\nonumber\\
0&=&\mu_{\Delta^0}-2\mu_{\Sigma^{*0}}+\mu_{\Xi^{*0}}
\end{eqnarray}
are valid to all orders in $1/N_c$ but first order in $m_s$.  Finally, one has
results 
\begin{eqnarray}
\mu_{\Delta^0}&=&0\nonumber\\
\mu_{\Delta^+}&=&3(\mu_p+\mu_n)\nonumber\\
\mu_{\Sigma^{*+}}-\mu_{\Sigma^{*-}}&=&3(-\mu_p+\mu_{\Sigma^-}+3\mu_{\Sigma^+}
-\mu_{\Xi^-}+2\mu_{\Xi^0})
\end{eqnarray}
which are valid to all orders in $m_s$ and first order in $1/N_c$.  Details of
these and other relations can be found in refs. \cite{lmw}. 

\medskip

Besides their masses and magnetic moments one of the most important 
properties of the hyperons is that they are unstable and decay via the
weak interaction.  It is this feature of hyperon physics to which we now turn.

\section{Hyperon Processes}

\subsection{Nonleptonic Hyperon Decay}
The dominant weak decay mode of the ${1\over 2}^+$ hyperons is 
the pionic channel $B\rightarrow B'\pi$, and the most general such decay
amplitude can be written in terms of an S-wave (parity-conserving)
amplitude---$A$---and P-wave (parity-violating) 
amplitude-- $B$---via\footnote{Note that our $\gamma_5$ is the 
negative of that defined
by Bjorken and Drell\cite{bj}.}
\begin{equation}
{\rm Amp}=\bar{u}(p')(A+B\gamma_5)u(p)
\end{equation}  
Such decays have been well-studied experimentally over the years and
both decay rates and asymmetries are well known.  From such measurements
one can determine both decay amplitudes and the resultant values are 
quoted in Table 5.
\begin{table}
\begin{center}
\begin{tabular}{ccccc}
Mode& $A^{exp}$&$A^{th}$& $B^{exp}$&$B^{th}$\\
$\Lambda^0\rightarrow p\pi^-$& 3.25&3.38&22.1&23.0\\
$\Lambda^0\rightarrow n\pi^0$&-2.37&-2.39&-15.8&-16.0\\
$\Sigma^+\rightarrow n\pi^+$&0.13&0.00&42.2&4.3\\
$\Sigma^+\rightarrow p\pi^0$&-3.27&-3.18&26.6&10.0\\
$\Sigma^-\rightarrow n\pi^-$&4.27&4.50&-1.44&-10.0\\
$\Xi^0\rightarrow \Lambda^0\pi^0$&3.43&3.14&-12.3&3.3\\
$\Xi^-\rightarrow\Lambda^0\pi^-$&-4.51&-4.45&16.6&-4.7
\end{tabular}
\caption{Decay amplitudes for Nonleptonic Hyperon Decay (in units of 
$10^{-7}$).  The theoretical amplitudes are the values arising from a
lowest order chiral fit.}
\end{center}
\end{table}

On the theoretical side there
remain two interesting and critical issues which have been with us since the
1960's---the origin of the $\Delta I=1/2$ rule and the S/P-wave 
problem\cite{dsm}:
\begin{itemize}

\item [i)]  The
former is the feature that $\Delta I=3/2$ amplitudes are suppressed with
respect to their $\Delta I=1/2$ counterparts by factors of the order of 
twenty or so.  This can be seen from the separation of the decay
amplitudes into $\Delta I={1\over 2}$ and $\Delta I={3\over 2}$
components---{\it cf.} Table 6.\footnote{Here the isospin amplitudes
ae defined via
\begin{equation}
\begin{array}{ll}
{\cal O}_\Lambda^-=\sqrt{2}{\cal O}_\Lambda^{(1)}-{\cal
O}_\Lambda^{(3)}&
{\cal O}_\Lambda^0=-{\cal O}_\Lambda^{(1)}-\sqrt{2}{\cal
O}_\Lambda^{(3)}\\
{\cal O}_\Xi^0=-{\cal O}_\Xi^{(1)}-\sqrt{2}{\cal O}_\Xi^{(3)}&
{\cal O}_\Xi^-=\sqrt{2}{\cal O}_\Xi^{(1)}-{\cal O}_\Xi^{(3)}\\
{\cal O}_\Sigma^-={\cal O}_\Sigma^{(1)}+{\cal O}_\Sigma^{(3)}&
{\cal O}_\Sigma^+={1\over 3}{\cal O}_\Sigma^{(1)}
-{2\over 3}{\cal O}_\Sigma^{(3)}+X_\Sigma\\
{\cal O}_\Sigma^0=-{\sqrt{2}\over 3}{\cal
O}_\Sigma^{(1)}+{2\sqrt{2}\over 3}{\cal O}_\Sigma^{(3)}+\sqrt{1\over
2}X_\Sigma&\quad 
\end{array}
\end{equation} 
Note that here the coefficient $X_\Sigma$ is of mixed isospin character.}
\begin{table}
\begin{center}
\begin{tabular}{ccc}
Mode& $(A_3/A_1)^{exp}$&$(B_3/B_1)^{exp}$\\
$\Lambda\rightarrow N\pi$& 0.014&0.006\\
$\Sigma\rightarrow N\pi$&-0.017&-0.047\\
$\Xi\rightarrow\Lambda\pi$&0.034&0.023
\end{tabular}
\caption{Experimental ratio of $\Delta I={3\over 2}$ and $\Delta I={1\over 2}$
components of the decay amplitude.}
\end{center}
\end{table}

This suppression is well-known in both hyperon as well as kaon
nonleptonic decay and, despite a large body of theoretical work, there
still exists no simple explanation for its validity.  Indeed, the 
lowest order weak nonleptonic $\Delta S=1$ Hamiltonian 
\begin{equation}
{\cal O}_W=\bar{d}\gamma_\mu(1+\gamma_5)u\bar{u}\gamma^\mu(1+\gamma_5)s
\end{equation}
can be written in terms of $\Delta I={1\over 2},{3\over 2}$ components as
\begin{equation}
{\cal O}_W={1\over 2}{\cal O}_1+{1\over 10}{\cal O}_2+{1\over 15}{\cal
O}_3+{2\over 3}{\cal O}_4
\end{equation}
with
\begin{eqnarray}
{\cal O}_1&=&H_A-H_B\nonumber\\
{\cal O}_2&=&H_A+H_B+2H_C+2H_D\nonumber\\
{\cal O}_3&=&H_A+H_B+2H_C-3H_D\nonumber\\
{\cal O}_4&=&H_A+H_B-H_C
\end{eqnarray}
and
\begin{eqnarray}
H_A&=&\bar{d}\gamma_\mu(1+\gamma_5)u\bar{u}\gamma^\mu(1+\gamma_5)s\nonumber\\
H_B&=&\bar{d}\gamma_\mu(1+\gamma_5)s\bar{u}\gamma^\mu(1+\gamma_5)u\nonumber\\
H_C&=&\bar{d}\gamma_\mu(1+\gamma_5)s\bar{d}\gamma^\mu(1+\gamma_5)d\nonumber\\
H_D&=&\bar{d}\gamma_\mu(1+\gamma_5)s\bar{s}\gamma^\mu(1+\gamma_5)s
\end{eqnarray}
Here ${\cal O}_{1,2}$ are octet operators carrying $\Delta I={1\over
2}$, while ${\cal O}_{3,4}$ are members of a 27-plet and carry $\Delta
I={1\over 2},{3\over 2}$ respectively.  Thus the $\Delta
I=1/2$ and $\Delta I=3/2$ components are comparable.   

Of course, there exist important strong interaction corrections to
these results.   For example, taking the gluon exchange corrections to $H_A$
at leading order we find
\begin{eqnarray}
H_A&\rightarrow& H_A-{g^2\over 16\pi^2}\ln\left({M_W^2\over
\mu_H^2}\right)(3H_B-H_A)\nonumber\\
H_B&\rightarrow& H_B-{g^2\over 16\pi^2}\ln\left({M_W^2\over
\mu_H^2}\right)(3H_A-H_B)
\end{eqnarray}
where $\mu_H$ represents a typical hadronic mass scale.  The operators
\begin{equation}
{\cal O}_\pm={1\over 2}({\cal O}_A\pm{\cal O}_B)
\end{equation}
are form-invariant---${\cal O}_\pm\rightarrow c_\pm{\cal
O}_\pm$---with
\begin{equation}
c_\pm=1+d_\pm{g^2\over 16\pi^2}\ln\left({M_W^2\over
\mu_H^2}\right),\quad{\rm with}\quad
d_+=-2,\,d_-=+4
\end{equation}
Using the renormalization group these corrections can be summed to
leading log, yielding
\begin{equation}
{c_\pm(\mu_H)\over c_\pm(M_W)}=1+d_\pm{g^2\over
16\pi^2}\ln\left({M_W^2\over \mu_H^2)}\right)
\end{equation}
Similarly the remaining operators can be treated, yielding
renormalization group coefficients\footnote{Note: for simplicity we have
omitted the contribution of penguin and certain other higher order operators.}
\begin{eqnarray}
c_1&=& 1.90\nonumber\\
c_2&=& 0.14\nonumber\\
c_3&=& 0.10\nonumber\\
c_4&=& 0.49
\end{eqnarray}

Such effects {\it can} bring about a $\Delta I=1/2$ enhancement
of a factor of three to four\cite{llog}, so we still need to account for
a factor of five to six.  In the case of the baryons this factor likely
arises from the validity of what is called the Pati-Woo
theorem\cite{pw}, which requires $<B'|{\cal O}_2|B>=<B'|{\cal O}_3|B>=
<B'|{\cal O}_4|B>=0$  for
the basic three-quark, color-antisymmetric component of the
wavefunction.  That such simple baryon-baryon matrix elements are relevant to
the problem of nonleptonic hyperon decay follows from the current
algebra-PCAC constraint
\begin{equation}
\lim_{q\rightarrow 0}<\pi^aB'|{\cal O}_i|B>={-i\over
F_\pi}<B'|[Q_5^a,{\cal O}_i]|B>={-i\over F_\pi}<B'|[I^a,{\cal O}_i]|B>
\end{equation}
where the replacement of the axial charge $Q_5^a$ by the vector charge
(isospin) $I^a$ follows from the $V+A$ character of the weak current.

For kaons the origin of the $\Delta I={1\over 2}$ enhancement is not
as transparent and appears to be associated with detailed
dynamical structure. However, this is still a subject under
study\cite{ddyn}.  Interestingly the
one piece of possible evidence for significant $\Delta I={1\over 2}$ rule
violation comes from a hyperon
reaction----hypernuclear decay\cite{hnuc}, 
as will be discussed below\cite{dub}.

\item [ii)]  The S/P-wave problem is not as well known but has been a 
longstanding difficulty to those of us theorists who try to calculate these
things.  The standard approach to such decays goes back to current algebra
days and expresses the S-wave (parity-violating) amplitude---$A$---as
a contact term---the baryon-baryon matrix element of the 
axial-charge-weak Hamiltonian commutator.  The corresponding P-wave 
(parity-conserving) amplitude---$B$---uses a simple pole model 
({\it cf.} Figure 3) with the the weak baryon to baryon
matrix element given by a fit to the S-wave sector.  Parity-violating 
$BB'$ matrix elements are neglected in accord with the Lee-Swift 
theorem\cite{ls}, which requires the vanishing of such terms in the 
limit of SU(3) symmetry.  
\begin{figure}
\centerline{\psfig{figure=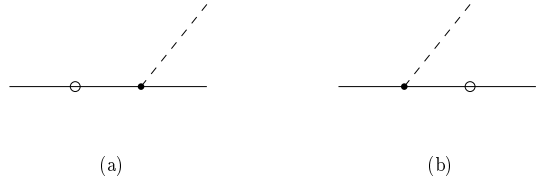,width=7.0cm}}
\caption{Pole diagrams used to calculated parity-conserving nonleptonic
hyperon decay.}
\end{figure}
With this procedure one can obtain a very good S-wave fit but finds 
P-wave amplitudes which are in very poor agreement with experiment.  
An example of such a fit is shown in Table 5.  (On 
the other hand, one can fit
the P-waves, in which case the S-wave predictions are very
unsatisfactory\cite{dghrv}).  Recently Skyrme as well
as QCD sum rule methods have also been applied to this problem 
with equal lack of 
success\cite{skr},\cite{qcd}.  

The original calculations had used simple 
relativistic effective interactions
to describe such processes.  However, contemporary treatments
generally involve the use of a chiral symmetry-based approach, in order to
keep track of terms of given chiral order.
One begins by representing the weak interaction in terms of a an
effective Lagrangian given in terms of an expansion in derivatives.
At lowest chiral order the form is unique
\begin{eqnarray}
{\cal L}_W^{pv}&=&
D{\rm tr}\left(\bar{B}\left\{u^\dagger\lambda_6u,B\right\}\right)\
+F{\rm
tr}\left(\bar{B}\left[u^\dagger\lambda_6u,B\right]\right)\nonumber\\
{\cal L}_W^{pc}&=&
D_5{\rm
tr}\left(\bar{B}\gamma_5\left\{u^\dagger\lambda_6u,B\right\}\right)
+F_5\left(\bar{B}\gamma_5\left[u^\dagger\lambda_6u,B\right]\right)
\end{eqnarray}
However, under a CP transformation 
\begin{equation}
B\rightarrow \left(i\gamma_2\bar{B}\right)^T,\qquad u\rightarrow
\left(u^\dagger\right)^T
\end{equation}
and it is seen that ${\cal L}_W^{pv}\rightarrow {\cal L}_W^{pv}$ but
${\cal L}_W^{pc}\rightarrow-{\cal L}_W^{pc}$ so we require
$D_5=F_5=0$, which is the Lee-Swift result\cite{ls}.  At leading
chiral order then the s-wave amplitude $A$ receives contributions
directly from ${\cal L}_W^{pv}$ and is ${\cal O}(p^0)$.  Likewise, the
p-wave amplitude $B$, which must arise from pole diagrams (Figure 3), is
also apparently ${\cal O}(p^0)$ by power counting.  
However, since the kinematic
structure $\bar{u}\gamma_5u\sim\vec{\sigma}\cdot\vec{q}$ the actual
behavior is $B\sim 1/\Delta m$ which is ${\cal O}(p^{-1})$.  

The S-wave amplitudes are then given in terms of just two 
parameters---$F,D$---via 
\begin{eqnarray}
A(\Lambda\rightarrow n\pi^0)&=&{i\over
2F_\pi}(3F+D)\nonumber\\
A(\Sigma^+\rightarrow p\pi^0)&=&{i\sqrt{6}\over
2F_\pi}(D-F)\nonumber\\
A(\Sigma^+\rightarrow
n\pi^+)&=&0\nonumber\\
A(\Xi^0\rightarrow\Lambda^0\pi^0)&=&{i\over 2F_\pi}(3F-D)\label{eq:df}
\end{eqnarray}
and a very good fit is provided by the choice
\begin{equation}
{F\over 2F_\pi}=0.92\times 10^{-7},\qquad {D\over F_\pi}=-0.42\label{eq:ch}
\end{equation}
as can be observed in Table 5.  (Note that this is one parameter less
than required by SU(3) symmetry alone.)  However, it is clear that the
P-wave predictions are very poor with this choice.

The first paper to examine chiral corrections to such predictions was that
of Bijnens et al. who calculated the leading nonanalytic behavior, 
{\it i.e.} leading log corrections, to nonleptonic hyperon decay within
a relativistic framework\cite{bij}.  However, when this was done the 
S-predictions no longer agreed with the data and the P-wave corrections 
were even larger.  Subsequently, Jenkins reexamined this system within 
the heavy baryon formalism.  As in ref. \cite{bij}, the calculation was
performed at leading log order, and she found that inclusion of decuplet
degrees of freedom in the loops significantly reduced the size of these
corrections, leading the restoration of experimental agreement in the case
of the S-wave amplitudes\cite{jen}.  However, the P-waves still were in
significant disagreement.  Perhaps in retrospect it is not so
surprising that the P-waves represent such a challenge, since they involve
significant cancellation in the sum of two pole contributions, which tends
to enhance both any SU(3) corrections and the effect of any leading log
modifications.  In any case it seems clear that some sort of additional 
contributions are required in order to achieve an understanding of these 
results, and an obvious tack is to perform a complete---not just leading
log---evaluation of corrections at one loop order. 

Recently Borasoy and Holstein (BH) took this approach and showed that
in a one loop heavy baryon chiral perturbative approach, it is certainly 
possible to fit the data by selection of appropriate values of the
various counterterms which arise\cite{bh0}.  However, there are 
many more counterterms
than data and one needs some guidance as to how to proceed.  In the BH
calculation, resonance saturation was used in order to decide which
counterterms to utilize, but not to determine their value.  Although a
reasonable fit was achieved, the {\it physics} of such a
result is obscure.  In fact twenty years ago a possible solution to
the S-/P-wave problem was proposed by Le Youaunc et al.\cite{you}, 
who suggested the use of SU(6) symmetry and the inclusion of
contributions from the $(70,1^-)$ intermediate states.  In this case,
since the parity of such resonant states is negative, such terms contribute 
to the parity-{\it violating} amplitude $A$ via pole diagrams, yielding
\begin{eqnarray}
\delta A(\Lambda^0\rightarrow n\pi^0)&=&
i9\sqrt{3}C{M_\Lambda-M_N\over F_\pi}\nonumber\\
\delta A(\Sigma^+\rightarrow p\pi^0)&=&
-i27\sqrt{2}C{M_\Sigma-M_N\over F_\pi}\nonumber\\
\delta A(\Sigma^+\rightarrow n\pi^+)&=&0\nonumber\\
\delta A(\Xi^0\rightarrow
\Lambda\pi^0)&=&-i12\sqrt{2}C{M_\Xi-M_\Lambda\over F_\pi}
\end{eqnarray}
where $C$ is a phenomenological coupling associated with these resonance
states, and we note that such
contributions vanish in the limit of SU(3) symmetry, as required by
the Lee-Swift theorem.  Although at first sight this approach might seem to
complicate matters by including an additional parameter, this is
actually a very welcome modification, since our previous analysis was in
contradiction to a simple constituent quark model calculation of the
baryon-baryon matrix elements which yields
\begin{equation}
F=-D={\sqrt{3}\over 2}G_F\sin\theta_c\cos\theta_c|\psi_0(0)|^2
\end{equation}
The wavefunction at the origin depends on the model ({\it e.g.} in the
MIT bag picture we find
$|\psi_0(0)|^2=1.15\times 10^{-2}$
GeV$^3$) but the $D/F$ ratio is model-independent and is quite
inconsistent with the value $-0.42$ found in Eq. \ref{eq:ch}.  Using
$C$ and $D=-0.85F$ as independent parameters, 
we find the very satisfactory fit 
shown in Table 7.  The numerical value---${F\over 2F_\pi}\simeq
10^{-7}$---is also in the right ballpark.  A full fit to s-waves and 
p-waves is also successful, as shown in the same table.  

Having observed the success of this resonant state insertion within the simple 
constituent quark model, it is natural to attempt a parallel approach 
from a chiral symmetry viewpoint.  Although a full one loop calculation 
has yet to be undertaken BH, in a preliminary calculation,
examined nonleptonic hyperon decay via inclusion of the lowest ${1\over 2}^-$ 
and ${1\over 2}^+$ resonant contributions\cite{bh1}.  (Note that they did
{\it not} include the entire complement of (70,$1^-$) states, which would
also include the decuplet states.)  A satisfactory 
picture of the nonleptonic decays was achieved in this approach, 
which reinforces the assertions of the Orsay group.  However, a full 
one loop calculation is still needed in order to firm up this
conclusion, and an important step in this direction was recently taken
by Borasoy and M\"{u}ller, who have evaluated the divergent components
of the counterterms\cite{bmu}.
\end{itemize}

\begin{table}
\begin{center}
\begin{tabular}{ccccccc}
Mode&$A^{exp}$& $A^{th1}$&$A^{th2}$&$B^{exp}$&$B^{th1}$&$B^{th2}$\\
$\Lambda^0\rightarrow p\pi^-$&3.25&2.0&2.17&22.1&24.5&23.4\\
$\Sigma^+\rightarrow n\pi^+$&0.13&0&-0.04&42.2&44.0&44.4\\
$\Sigma^-\rightarrow n\pi^-$&4.27&4.42&5.33&-1.44&0.9&-1.61\\
$\Xi^-\rightarrow \Lambda\pi^-$&-4.51&-4.70&-4.14&16.6&19.2&14.7
\end{tabular}
\caption{Calculated vs. experimental 
hyperon decay amplitudes in a simple SU(6) picture 
($th1$) and in a chiral model with resonance contributions ($th2$).  All
amplitudes should be multipled by the factor $10^{-7}$.}
\end{center}
\end{table} 
\noindent It should perhaps be emphasized that both in the case of the 
$\Delta I={1\over 2}$ Rule and of the S-/P-wave problem, we do {\it not} 
require more and better data.  The issues are already clear.  What we need is 
more and better theory!

Where one {\it does} need data involves the possibility of testing 
the standard model prediction of CP violation, which predicts the presence 
of various asymmetries
in the comparision of hyperon and antihyperon nonleptonic decays\cite{dcp}.  
The basic
idea is that one can write the decay amplitudes in the form
\begin{equation}
A=|A|\exp i(\delta_S+\phi_S),\quad B=|B|\exp i(\delta_P+i\phi_P)
\end{equation} 
where $\delta_S,\delta_P$ are the strong S,P-wave phase shifts at the 
decay energy of the mode being considered and $\phi_S,\phi_P$ are CP-violating
phases which are expected to be of order $10^{-4}$ or so in standard model
CP-violation.  One can detect such phases by comparing hyperon and antihyperon
decay parameters.  Unfortunately nature is somewhat perverse here in that
the larger the size of the expected effect, the more difficult the 
experiment.  For example, the asymmetry in the overall decay rate, which is
the easiest to measure, has the form\footnote{Here $B^r$ indicates a reduced
amplitude---$B^r= B(E'-M_{B'})/(E'+M_{B'})$.}
\begin{eqnarray}
C_\Gamma&=&{\Gamma-\bar{\Gamma}\over \Gamma+\bar{\Gamma}}\nonumber\\
&\sim& 
\left(-2(A_1A_3\sin(\delta_S^1
-\delta_S^3)\sin(\phi_S^1-\phi_S^3)\right.\nonumber\\
&+&\left.B_1^rB_3^r\sin(\delta_P^1-\delta_P^3)\sin(
\phi_P^1-\phi_P^3)\right)\nonumber\\
&/& |A_1|^2+|B_1^r|^2
\end{eqnarray}
where the subscripts, superscripts 1,3 
indicate the $\Delta I={1\over 2},{3\over 2}$ 
component of the amplitude.  We see then that there is indeed sensitivity to
the CP-violating phases but that it is multiplicatively 
suppressed by {\it both} the  
strong interaction phases ($\delta\sim 0.1$) as well as by the $\Delta I=
{3\over 2}$ suppression $A_3/A_1\sim B_3/B_1\sim 1/20$.  Thus we 
find $C\sim\phi/100\sim 10^{-6}$
which is much too small to expect to measure in present generation experiments.

More sanguine, but still not optimal, is a comparison of the asymmetry
parameters $\alpha$, defined via
\begin{equation}
W(\theta)\sim1+\alpha\vec{P}_B\cdot\hat{p}_{B'}
\end{equation}
In this case, one finds
\begin{eqnarray}
C_\alpha&=&{\alpha+\bar{\alpha}\over \alpha-\bar{\alpha}}
=-\sin(\phi_S^1-\phi_P^1)
\sin(\delta_S^1-\delta_P^1)\nonumber\\
&\sim& 0.1\phi\sim 4\times 10^{-4}
\end{eqnarray} 
which is still extremely challenging.

Finally, the largest signal can be found in the combination
\begin{equation}
C_\beta={\beta+\bar{\beta}\over \beta-\bar{\beta}}=\cot(\delta_S^1
-\delta_P^1)\sin(\phi_S^1-\phi_P^1)\sim\phi
\end{equation}
Here, however, the parameter $\beta$ is defined via the general expression
for the final state baryon polarization
\begin{eqnarray}
<\vec{P}_{B'}>&=&{1\over W(\theta)}\left((\alpha+\vec{P}_B\cdot\hat{p}_{B'})
\hat{p}_{B'}\right.\nonumber\\
&+&\left.\beta\vec{P}_B\times\hat{p}_{B'}
+\gamma(\hat{p}_{B'}
\times(\vec{P}_B\times\hat{p}_{B'}))\right)\nonumber\\
\quad
\end{eqnarray}
and although, compared to $C_\Gamma$ and $C_\alpha$, 
the size of the effect is largest---$C_\beta\sim 10^{-3}$---this
measurement seems out of the question experimentally.  

Despite the small size of these effects, the connection with standard model
CP violation and the possibility of finding larger effects due to new
physics demands a no-holds-barred effort to measure these parameters.

\subsection{Hypernuclear Decay}

Above we mentioned the possibility of $\Delta I={1\over 2}$ Rule violation
in hypernuclear decay.  Since this represents an interesting and challenging
area of hyperon physics, we present here a simple and concise 
introduction to some of the issues.  However, chiral symmetry has 
not yet been applied
to this realm, so our discussion will utilize conventional analysis.

A hypernucleus is a nuclear system in which one of the neutrons has been 
replaced by a $\Lambda$.  As outlined above, the dominant weak decay mode of 
a free $\Lambda$ is the pionic decay $\Lambda\rightarrow N\pi$.  One 
might anticipate then that in a hypernucleus, the decay rate is somewhat
modified by the nuclear medium but that essentially the same physics is
relevant.  This is not the case, however.  Indeed since the 
recoiling nucleon 
in the free decay has a momentum of only 100 MeV or so, while the Fermi
momentum of a typical nucleus is $k_F\sim 270$ MeV, the pionic decay of
the $\Lambda$ is strongly Pauli blocked\cite{chs}.  Instead, another 
mechanism, nonmesonic weak decay, takes 
place---$\Lambda+p\rightarrow n+p$ or $\Lambda+n\rightarrow n+n$.
In this case the recoiling nucleons carry momentum $p\sim 400$ MeV 
$>>k_F$, so this becomes the dominant decay channel.  
Such hypernuclear decays have been studied in recent years and one now has
a reasonable data base on various light and heavy nuclear systems, such as
${}_\Lambda^5He,\,\,{}_\Lambda^{11}B,\,\,{}_\Lambda^{12}C$, 
{\it etc.}  In general one
finds that the hypernuclear decay rate is roughly comparable to that for 
free $\Lambda$ decay and that the rates for neutron stimulated ($n\Lambda$)
and proton-stimulated ($p\Lambda$) decay are about the same\cite{exp}.   

On the theoretical side there has also been a good deal of work, much of it
based on a meson exchange picture of the effective interaction\cite{dfh}.  
The obvious meson of choice is the pion since it has the longest range and,  
in addition, the free $\Lambda$ decays via pion emission so
the weak couplings are known.  However, a realistic description 
must also include
heavier (short-distance) exchanges of $K,\eta,\rho,K^*,\omega,\ldots$ 
in order to be believable.
This is also demanded phenomenologically.  Indeed in a a pion-only 
exchange picture the effective weak interaction is of the form
\begin{equation}
{\cal H}_w\sim g\bar{N}\vec{\tau}N\cdot\bar{N}\vec{\tau}\left(
\begin{array}{l}
0\\1\end{array}\right)\Lambda
\end{equation}
Then $\Lambda n\rightarrow nn\sim g$ but $\Lambda p\rightarrow np
\sim(-1-\sqrt{2}^2)g=-3g$ so that p-stimulated decay would be
strongly dominant over its
n-stimulated counterpart, in contradistinction to experiment\cite{dfh}.  
Inclusion of
heavier meson exchange (especially K-exchange) tends to help in this regard.
However, no reasonable picture has been able to achieve the comparable values
of p-stimulated and n-stimulated decay which are seen experimentally.  
On the other hand, all of these comprehensive
meson-exchange analyses have, for simplicity, {\it assumed} the validity 
of the $\Delta I={1\over 2}$ Rule both for pion exchange, where it has
been experimentally verified, as well as for heavy meson exchange, where no
data exists.  Some theorists have postulated the existence of 
possible $\Delta I={1\over 2}$ Rule violating four fermion interactions
which may be
relevant and it becomes an experimental issue whether such violations are
present or not\cite{ms1}.  The problem is that the data base is small and the 
uncertainties large so that substantial violation must be found in order to
be believable.

At the present time there do exist suggestive results which may 
possibly indicate the
presence of significant $\Delta I={3\over 2}$ effects in hypernuclear 
decay\cite{d32}.  This comes from analysis of nonmesonic hypernuclear 
decay in light 
systems---${_\Lambda^4H,{}_\Lambda^5He,{}_\Lambda^4He}$---wherein one can
reasonably assume that the interaction involves only S-waves.  The possible
$\Lambda N$ transitions can then be categorized via
\begin{equation}
\begin{array}{ccc}
{}^SL_J(initial)&{}^SL_J(final)& I_f\\
{}^1S_0&{}^1S_0&1\\
{}^1S_0&{}^3P_0&1\\
{}^3S_1&{}^3S_1&0\\
{}^3S_1&{}^1P_1&0\\
{}^3S_1&{}^3P_1&1\\
{}^3S_1&{}^3D_1&0
\end{array}
\end{equation}
and, in a semiclassical approach to such decays, one can write
\begin{equation}
\Gamma_{NM}({}_\Lambda^{A+1}Z)={N\bar{R}_n+Z\bar{R}_p\over A}\rho_A
\end{equation}
where $\rho_A$ is the average nucleon density at the location of the $\Lambda$
particle and $\bar{R}_n,\bar{R}_p$ denote the spin-averaged rate for the 
neutron-induced, proton-induced decay respectively.  Defining\cite{dal}
\begin{eqnarray}
R_{n0}&=&R_n({}^1S_0)+R_n({}^3P_0)\nonumber\\
R_{p0}&=&R_p({}^1S_0)+R_p({}^3P_0)\nonumber\\
R_{n1}&=&R_n({}^3P_1)\nonumber\\
R_{p1}&=&R_p({}^3S_1)+R_p({}^1P_1)+R_p({}^3P_1)+R_p({}^3D_1)
\end{eqnarray} 
we can write the light hypernuclear decay rates as
\begin{eqnarray}
\Gamma_{NM}({}_\Lambda^3H)&=&{\rho_3\over 8}(3R_{n0}+R_{n1}+3R_{p0}+R_{p1})
\nonumber\\
\Gamma_{NM}({}_\Lambda^4H)&=&{\rho_4\over 6}(R_{n0}+3R_{n1}+2R_{p0})\nonumber\\
\Gamma_{NM}({}_\Lambda^4He)&=&{\rho_4\over 6}(2R_{n0}+R_{p0}+3R_{p1})
\nonumber\\
\Gamma_{NM}({}_\Lambda^5He)&=&{\rho_5\over 8}(R_{n0}+3R_{n1}+R_{p0}+3R_{p1})
\end{eqnarray}
If the $\Delta I={1\over 2}$ Rule obtains then we have
\begin{equation}
R_n({}^1S_0)=2R_p({}^1S_0),\,\,\,R_n({}^3P_0)=2R_p(^3P_0),\,\,\,R_n({}^3P_1)
=2R_p({}^3P_1)
\end{equation}
Consequently, 
\begin{equation}
{R_{n1}\over R_{p1}}\leq{R_{n0}\over R_{p0}}=2
\end{equation}
which should be compared with the experimental results\cite{d32}
\begin{equation}
{R_{n0}\over R_{p0}}=0.6^{+1.3}_{-0.6},\qquad{R_{n1}\over R_{p1}}
=1.0^{+1.1}_{-1.0}
\end{equation}
Obviously the statistics do not permit the drawing of a definitive conclusion
at this time, but these suggestive results cry out for higher precision
experiments as well as improved theory.

\subsection{Nonleptonic Radiative Decay}
Another longstanding thorn in the side of theorists attempting to understand
weak decays of hyperons is the nonleptonic radiative mode $B\rightarrow
B'\gamma$\cite{zrev}.  In this case one can write 
the most general decay amplitude as 
\begin{eqnarray}
{\rm Amp}&=&{-ie\over M_B+M_{B'}}\epsilon^{*\mu}q^\nu\nonumber\\
&\times&\bar{u}(p')\sigma_{\mu\nu}(C+D\gamma_5)u(p)
\end{eqnarray}
where $C$ is the magnetic dipole (parity-conserving) amplitude and $D$ is its
(parity-violating) electric dipole counterpart.  There are then two primary 
quantities of interest in the analysis of such decays---the decay rate
\begin{equation}
\Gamma_{B\rightarrow B'\gamma}={\alpha\over 2}\left(M_B-M_{B'}\over
M_B\right)^3(M_B+M_{B'})(|C|^2+|D|^2)
\end{equation}
and the asymmetry parameter
\begin{equation}
A_{B\rightarrow B'\gamma}=-{2{\rm Re}C^*D\over |C|^2+|D|^2}
\end{equation}
where the angular distribution is given (in the cm frame) by 
\begin{equation}
{d\Gamma_{B\rightarrow B'\gamma}\over d\Omega}={\Gamma_{B\rightarrow
B'\gamma}\over 4\pi}\left(1+A_{B\rightarrow
B'\gamma}\vec{\sigma}_{B}\cdot\hat{p}_{B'}\right)\, .
\end{equation}

Such decays have been the subject of a great deal of experimental and 
theoretical scrutiny for well over three decades.  Present results are
summarized in Table 8.
\begin{table}
\begin{center}
\begin{tabular}{ccc}
Mode & Branching Fraction ($\times 10^{-3}$) & Asymmetry \\
$\Sigma^+\rightarrow p\gamma$&$1.23\pm 0.06$ &$ -0.76\pm 0.08$\\
$\Lambda\rightarrow n\gamma$ & $1.63\pm 0.14$ & \\
$\Xi^0\rightarrow\Lambda\gamma$&$1.06\pm 0.16$&$ 0.43\pm 0.44$\\
$\Xi^0\rightarrow\Sigma^0\gamma$&$3.56\pm 0.43$&$0.20\pm 0.32$\\
$\Xi^-\rightarrow\Sigma^-\gamma$&$0.128\pm 0.023$&$1.0\pm 1.3$\\
$\Omega^-\rightarrow\Xi^-\gamma$& $<0.46$& 
\end{tabular}
\caption{Hyperon Radiative Decay Measurements}
\end{center}
\end{table}
However, a basic theoretical understanding 
remains elusive.  The fundamental difficulty is associated with what
has come to be called ``Hara's Theorem'', which requires that
in the SU(3) limit the parity violating decay amplitude must vanish
for CP-conserving transitions
between states of a common U-spin multiplet---{\it i.e.} $\Sigma^+\rightarrow
p\gamma$ and $\Xi^-\rightarrow \Sigma^-\gamma$\cite{hara}.  The proof
here is very much analogous to that which requires the vanishing of
the axial tensor form factor in nuclear beta decay between members of
a common isotopic spin multiplet\cite{ht}.  One performs a
U-spin rotation $\exp(-i\pi U_2)$ ({\it i.e.}, in quark language
interchanges $s,d$ quarks) on the matrix element 
\begin{equation}
\int d^4xe^{iq\cdot x}<B'|T(V_\mu^{em}(x){\cal H}_w(0)|B>
\end{equation}
responsible for the radiative weak decays.  Since both the
electromagnetic current and weak Hamiltonian are invariant under such
a transformation, we see that the amplitude for $B\rightarrow
B'\gamma$ and $B'\rightarrow B\gamma$ must be identical when $B,B'$
are members of a common U-spin multiplet ({\i.e.} $\Sigma^+,p$ or 
$\Xi^-,\Sigma^-$).  On the other hand, CP invariance requires the form
of the effective interaction to be
\begin{equation}
{\cal H}_{eff}=-iF^{\mu\nu}\left[\bar{\psi}_{B'}{\sigma_{\mu\nu}}
(a+b\gamma_5)\psi_B+\psi_{B}\sigma_{\mu\nu}(a-b\gamma_5)\psi_{B'}\right]
\end{equation}
Only the parity-conserving interaction---$a$---is consistent with
the U-spin stricture and consequently we require $b=0$, which 
is Hara's theorem.
  
Of course, this theorem is only strictly true in the SU(3) limit.  
Nevertheless, since only $\sim 20\%$ SU(3) breaking effects are expected, 
one anticipates a relatively small photon
asymmety parameter for such decays.  In 1969 the
first measurement of such a quantity---based upon 61 events---was
reported, yielding the surprisingly large value\cite{gersh}
\begin{equation}
A_{\Sigma^+\rightarrow p\gamma}=-1.03^{+0.52}_{-0.42}
\end{equation}
and since that time the measurement (now based on nearly 36,000
events) has been refined, but remains large and negative, as shown in
Table 8.\cite{pdg}  For these thirty years theorists have been
struggling to explain this result. 

\begin{figure}
\centerline{\psfig{figure=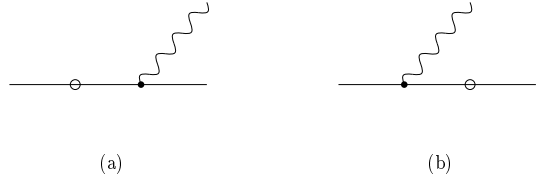,width=7cm}}
\caption{Pole diagrams used to calculate radiative hyperon decay.}
\end{figure}

To leading chiral order the amplitude is given by the simple pole 
diagrams shown in Figure 4, with
the weak baryon-baryon matrix elements being those
determined in the pionic decay analysis discussed above.  
Such matrix elements---$<B'|{\cal H}_w|B>$---are generally taken to be
purely parity conserving since the Lee-Swift theorem asserts that all weak
parity-violating baryon-baryon
matrix elements must vanish in the SU(3) limit and it is in this fashion
that Hara's theorem is manifested in a pole model\cite{ls}. 
The results of
such an analysis are generally satisfactory (though only in an order of
magnitude sense) for the branching ratio,
but, since D=0, the
asymmetry parameter is predicted to vanish for {\it all} radiative hyperon
transitions, in contradistinction to experiment.  It is possible to utilize
a simple constituent quark model to estimate SU(3) breaking
corrections.\cite{gh}  However, such calculations yield only $D/C \leq 0.05$
and certainly do not appear to be large enough to explain 
the experimental findings.

In a chiral approach, one finds similar problems.  At lowest order in
the chiral expansion we have
\begin{eqnarray}
C_{n\Sigma^0}&=&{8\sqrt{2}\over 3}\left({M_\Sigma+M_N\over
M_\Sigma-M_N}\right)Dd_\mu\nonumber\\
C_{n\Lambda}&=&{8\over 3}\sqrt{2\over 3}\left({M_\Lambda+M_N
\over M_\Lambda-M_N}\right)Dd_\mu\nonumber\\
C_{\Sigma^0\Xi^0}&=&{8\sqrt{2}\over 3}\left({M_\Xi+M_\Sigma\over 
M_\Xi-M_\Sigma}\right)Dd_\mu\nonumber\\
C_{\Lambda\Xi^0}&=&{8\over 3}\sqrt{2\over
3}\left({M_\Xi+M_\Lambda\over
M_\Xi-M_\Lambda}\right)Dd_\mu\nonumber\\
C_{p\Sigma^+}&=&C_{\Sigma^-\Xi^-}=0
\end{eqnarray}    
where here $d_\mu$ is the d-coupling coefficient in a lowest order
SU(3) fit to the magnetic moments, D is the weak SU(3) coupling defined
in Eq. \ref{eq:df}, and $D_{B'B}=0$ for all such decays.  
We observe the
vanishing of the entire decay rate for the U-spin doublet decays and
of the asymmetry for the remaining modes, both
predictions being in serious disagreement with experiment.  

As in the case of the S-/P-wave puzzle for ordinary nonleptonic hyperon
decay, what seems clearly required here is the
inclusion of additional physics, and one may optimistically expect 
that the same
mechanism which solves the S/P-wave problem may also work here.  
It is reassuring in this regard that the work two decades ago by
the Orsay group proposing the inclusion of $(70,1^-)$ states also seemed
to solve the Hara's theorem problem.   In the case of a contemporary chiral
approach, a one-loop calculation has been performed but has not led to a 
resolution, although somewhat higher values of the asymmetry can be 
accomodated\cite{neu}.  However, Borasoy and Holstein have, just as in 
their analysis of pionic hyperon decay, calculated within a chiral
framework the contribution of the lowest ${1\over 2}^-$ and ${1\over 2}^+$ 
resonances to radiative hyperon decay\cite{bh2}. One finds, for example,
\begin{eqnarray}
C_{p\Sigma^+}&=&{4(M_\Sigma+M_N)\over 3}\left({1\over
3}\ell_d+\ell_f\right)(d^*-f^*)\left({1\over M_\Sigma-M_{B*}}
+{1\over M_N-M_{B*}}\right)\nonumber\\
D_{p\Sigma^+}&=&4\left({(M_\Sigma-M_N)(M_\Sigma+M_N)\over 
(M_\Sigma-M_{B*})(M_N-M_{B*})}\right)\left({1\over 3}r_d+r_f\right)(w_d-w_f)
\end{eqnarray}
where here $\ell_d,\ell_f$ and $r_d,r_f$ represent SU(3) symmetric,
antisymmetric electromagnetic couplings connecting the hyperons 
to ${1\over 2}^-$ and ${1\over 2}^+$ resonant states respectively,
while $d^*,f^*$ and $w_d,w_f$ represent the corresponding weak
couplings to such resonances.  Note that in this picture, the 
form of the parity-violating
coupling $D$ explicitly satisfies Hara's theorem, as required, while
while in addition, for U-spin siblings, the parity-conserving coupling 
$C$ vanishes unless coupling to resonances is included.  The
size of the electromagnetic couplings $\ell_d\,\ell_f$ can be taken from 
experiment, while the weak couplings $w_f,w_d,f^*,d^*$ are determined from the 
previous fit to pionic nonleptonic hyperon decay, yielding a no-free-parameter 
prediction to this order.  

What is found is that relatively large negative values of the
asymmetry seem to arise naturally in such a picture.  That is not to
say that perfect agreement is obtained, but neither is this a complete
calculation.  In the ideal case, a complete one-loop 
calculation---{\it including low lying resonaces}---should be done 
for {\it both} ordinary and radiative hyperon decay
and a fit to both should be performed.  This lies in the future.  
However, in the absence of
such a program, the present results are encouraging and
suggest that we are on the right track.

Before closing this section, it should be noted, however, that the field 
is littered with red herrings, which have tended to confuse expert and 
nonexpert alike.  These have appeared in the guise of supposed Hara's 
theorem-{\it violating} models which claim to explain the radiative 
asymmetry results.
An example is a calculation by Kamal and Riazuddin, who
evaluated the amplitude for radiative decay in a simple quark model 
with W-exchange\cite{kr}.
In the static limit, these authors determined that there exists a
parity-violating contribution to the decay
\begin{equation}
{\cal H}_{pv}\sim\vec{\epsilon}\cdot(u^\dagger\vec{\sigma}s)
\times(d^\dagger\vec{\sigma}u)
\end{equation}
which is non-vanishing even in the limit of SU(3)-symmetry!  However,
gauge invariance is violated in this approach.  Another gauge
invariance violating procedure invokes vector dominance.  
However, as emphasized by Dmitrasinovic, such apparent violations are 
illusory\cite{dmi}---such results are calculated in a simplistic 
models and do {\it not} satisfy the
usual field theoretic constraints of crossing symmetry, gauge
invariance, {\it etc.}  It is well-known in nuclear physics that when one
calculates an electromagnetic matrix element in simple impulse
approximation, the result generally violates gauge invariance unless many-body
effects associated with meson exchange are included, which is often
accomplished by the use of Siegert's theorem.  The same is true here---the
quark calculation is the particle physics analog of the impulse
approximation and, in such an elementary calculation, it is no surprise
that fundamental symmetries such as gauge invariance (and Hara's
theorem) become violated.  This is simply {\it not} a consistent field
theoretic calculation and must therefore be discounted.
  
In order, however, to confirm the validity of this or any model what 
will be required is a set of measurements of {\it both} rates and 
asymmetries for such decays.  In this regard, it should be noted 
that theoretically one
expects all asymmetries to be negative in any realistic
model\cite{pz}---it would be very difficult
to accomodate any large positive asymmetry.  Thus the present particle data
group listing\cite{pdg} 
\begin{equation}
A_\gamma(\Xi^0\rightarrow\Lambda\gamma)=+0.43\pm0.44
\end{equation}      
strongly suggests the need for a new measurement.

\subsection{Hyperon Semileptonic Decay}
A mode that theory does well in predicting (in fact some would say 
{\it too} well) is that of hyperon beta decay---$B\rightarrow 
B'\ell\bar{\nu}_\ell$,
where $\ell$ is either an electron or a muon.  Since this is a semileptonic
weak interaction, the decays are described in general by matrix elements of
the weak vector, axial-vector currents
\begin{eqnarray}
<B'|V_\mu|B>&=&\bar{u}(p')(f_1\gamma_\mu+{-if_2\over M_B+M_{B'}}
\sigma_{\mu\nu}q^\nu\nonumber\\
&+&{f_3\over M_B+M_{B'}}q_\mu)u(p)\nonumber\\
<B'|A_\mu|B>&=&\bar{u}(p')(g_1\gamma_\mu+{-ig_2\over M_B+M_{B'}}
\sigma_{\mu\nu}q^\nu\nonumber\\
&+&{g_3\over M_B+M_{B'}}q_\mu )\gamma_5u(p)\nonumber\\
\quad
\end{eqnarray}
Here the dominant terms are the vector, axial couplings $f_1,g_1$ and the
standard approach is simple Cabibbo theory, wherein one fits the $g_1$
in terms of SU(3) coefficients $f_A,d_A$ and $f_1$ using
CVC and simple F-type coupling, since the vector interaction is protected
to first order from renormalization by the Ademollo-Gatto theorem.\cite{ag}
(Note that the lowest order chiral invariance predictions agree here with
the traditional SU(3) symmetry approach.)
\begin{table}
\begin{center}
\begin{tabular}{ccc}
mode&vector&axial\\
pn&$1$&$f_A+d_A$\\
$\Lambda\Sigma^-$&0&${2\over \sqrt{6}}d_A$\\
$p\Lambda$&-$\sqrt{3\over 2}$&$-{1\over \sqrt{6}}(d_A+3f_A)$\\
$\Lambda\Xi^-$&$\sqrt{3\over 2}$&$-{1\over \sqrt{6}}(d_A-3f_A)$\\
$n\Sigma^-$&-1&$d_A-f_A$\\
$\Sigma^0\Xi^-$&$\sqrt{1\over 2}$&$\sqrt{1\over 2}(d_A+f_A)$
\end{tabular}
\caption{Lowest order vector and axial coupling constants for hyperon
beta decay.}
\end{center}
\end{table}
When this is done\footnote{Note that it is essential to include radiative
corrections here.} one finds in general a 
very satisfactory fit---$\chi^2/d.o.f\sim 3$---which can be made even better 
by inclusion of simple quark model SU(3) breaking effects---$\chi^2/d.o.f.
\sim 1$\cite{dhk}\cite{ra}.  An output of such a fit is a 
value of the KM mixing parameter
$V_{us}=0.220\pm 0.003$, which is in good agreement with the value
$V_{us}=0.2196\pm 0.0023$ measured in $K_{e3}$ decay.  
However, differing assumptions about SU(3) 
breaking will lead to slightly modified numbers.  

The importance of such a measurement of $V_{us}$ has to do with its use as
an input to a test of the standard model via the unitarity prediction
\begin{equation}
|V_{ud}|^2+|V_{us}|^2+|V_{ub}|^2=1
\end{equation} 
From an analysis of B-decay one obtains $|V_{ub}|\sim 0.003$, which when 
squared leads to a negligible contribution to the unitarity sum.  So the
dominant effect comes from $V_{ud}$, which is measured via $0^+-0^+$
superallowed nuclear beta decay---
\begin{equation}
V_{ud}^2={2\pi^3\ln 2m_e^{-5}\over 2G_F^2(1+\Delta_R^V)\bar{\cal F}t}
\end{equation}
Here $\Delta_R^V=2.40\pm 0.08\%$ is the radiative correction and 
$\bar{\cal F}t= 3072.3\pm 0.9$ sec. is the mean (modified) ft-value
for such decays.  Of course, there exist important issues in the analysis
of such ft-values including the importance of isotopic spin breaking
effects and of possible Z-dependence omitted from the radiative corrections, 
but if one takes the above-quoted number as being correct then\cite{htow}
\begin{equation}
\begin{array}{c}
V_{ud}=0.9740\pm0.0005\\
{\rm and}\\
|V_{ud}|^2+|V_{us}|^2+|V_{ub}|^2=
0.9968\pm 0.0014
\end{array}
\end{equation}
which indicates a possible violation of unitarity.  If correct, this would
suggest the existence of non-standard-model physics, but clearly additional
work, both theoretical and experimental, is needed before drawing this
conclusion.

A second interesting implication of the hyperon beta decay analysis has 
to do with the
strange quark content of the nucleon.  The issue here starts with the 
observation that the axial matrix element in neutron beta decay can be written
in terms of the integrated u,d,s helicity content of the nucleon via
\footnote{Specifically, $\Delta q$ represents the net helicity of the 
quark flavor $q$ along the direction of the proton spin in the infinite 
momentum frame---
\begin{equation}
\Delta q=\int_0^1 dx\Delta q(x)=\int_0^1dx(q^\uparrow(x)+\bar{q}^\uparrow(x)
-q^\downarrow(x)-\bar{q}^\downarrow(x))
\end{equation}}
\begin{equation}
g_A(np)=\Delta u-\Delta d
\end{equation}
while from hyperon beta decay we have the related result
\begin{equation}
g_A(np)\left(3f_A-d_A\over f_A+d_A\right)=\Delta u+\Delta d-2\Delta s
\end{equation}
Thus far, there are only two equations but three unknowns, so we require
an additional constraint.  One possibility is to assume that $\Delta s=0$, 
{\it i.e.} to neglect the possibility of strangeness quark content of the 
nucleon.  This is usually called the Ellis-Jaffe sum rule\cite{ejaf} 
and, taking $g_A=1.267$ and the value $f_A/d_A=0.58$ from the simple hyperon
decay fit, yields the
prediction for the integrated
spin structure function $g_1(x,Q^2)$ measured in deep 
inelastic longitudinally polarized electron scattering
\begin{eqnarray}
\Gamma_1^p(Q^2)&=&\int_0^1dxg_1(x,Q^2)=\nonumber\\
&=&{1\over 2}\left({4\over 9}\Delta u+
{1\over 9}\Delta d +{1\over 9}\Delta s\right)(1+\delta_{QCD})\nonumber\\
&=&({\rm Ellis-Jaffe})\,\,\,0.171\pm 0.006\,\,{\rm at}\,\,Q^2=10\,\,{\rm GeV^2}
\end{eqnarray}
On the other hand the experimentally measured value 
\begin{equation}
\Gamma_1(Q^2)^{\rm exp}=0.120\pm0.005\pm 0.006
\end{equation}
is quite different and yields a nonzero value for the strange quark 
content $\Delta s=-0.11\pm 0.01$, 
which is perhaps not unreasonable.  But it should be noted that a relatively 
small shift to the value $f_A/d_A\simeq 0.5$ could change this result to 
$\Delta s=0$.  (As an aside it should be noted that these values for the
$\Delta q$ are also used in order to calculate the 
fraction of nucleon spin which arises from valence quarks.  Specifically,
using the above inputs we have
\begin{equation}
\Delta u+\Delta d +\Delta s=0.25\pm 0.04.
\end{equation}
{\it i.e.} most of the nucleon spin comes from its nonvalence components.)

Higher order chiral corrections to the lowest order SU(3) (chiral) 
predictions have also been investigated.  The first such calculation 
was of the leading nonanalytic loop corrections
by Bijnens, Sonoda, and Wise and yielded results in the form\cite{bsw}
\begin{equation}
g_1(ij)=\sqrt{Z_iZ_j}\left[\alpha_{ij}+{1\over 16\pi^2F_\pi^2}\sum_k
\beta^k_{ij}m_k^2\ln{m_k^2\over \mu^2}\right]
\end{equation}
where the leading order coefficients $\alpha_{ij}$ are given in Table 9, the 
axial renormalization couplings $\beta^k_{ij}$ can be found in ref. \cite{bsw}
and $Z_i$ are wavefunction renormalization factors whose leading nonanalytic
form is
\begin{equation}
Z_i=1-{1\over 16\pi^2F_\pi^2}\kappa_i^jm_j^2\ln{m_j^2\over \mu^2}
\end{equation}
When such corrections are included, however, the fit becomes {\it much}
worse---$\chi^2$/d.o.f.=2.9.  Of course, such problems can be solved by
inclusion of higher order counterterms, but this program has yet to be
carried out and in any case one wonders yet again about the phenomenological
utility of the chiral expansion.  The use of cutoff regularization
can help.  In this case the dipole-modifed integral includes a quadratic
divergence which can be absorbed into renormalized values of the axial
couplings via
\begin{eqnarray}
d_A^r&=&d_A-{3\over 2}{\Lambda^2\over 16\pi^2F_\pi^2}d_A(3d_A^2+5f_A^2+1)
\nonumber\\
f_A^r&=&f_A-{1\over 6}{\Lambda^2\over 16\pi^2F_\pi^2}f_A(25d_A^2+63f_A^2+9)
\end{eqnarray}  
and the cutoff-reularized integral brings the chiral corrections
under reasonable control, as found in the case of the masses.

An alternative approach is that of evaluating the leading HB$\chi$pt 
loop corrections with a $1/N_c$ expansion while retaining the nonzero 
$N,\Delta$ mass difference which obtains experimentally\cite{mhjm}.
It is seen in such a calculation that the loop corrections to the axial
currents retain a substantial cancellation which is due to the feature that
the octet and decuplet baryons share a flavor-spin symmetry in the large
$N_c$ limit.  However, the significance of this feature as to data analysis
has yet to be explored.

What is needed in the case of hyperon beta decay is good set of data including
rates {\it and} asymmetries, both in order to produce a possibly improved
value of $V_{us}$ and also to study the interesting issue of SU(3) breaking
effects, which {\it must} be present, but whose effects seem somehow
to be hidden.  A related focus of such studies should be the
examination of higher
order---recoil---form factors such as weak magnetism ($f_2$) and the axial
tensor ($g_2$).  In the latter case, Weinberg showed that in the standard
quark model $G=C\exp (i\pi I_2)$-invariance requires $g_2=0$ in neutron
beta decay $n\rightarrow pe^-\bar{\nu}_e$\cite{scc}.  
(This result usually is called
the stricture arising from ``no second class currents'' and has been
verified to reasonable precision by correlation experiments in
nuclear beta decay\cite{min}.)  In the SU(3)
limit one can use V-spin invariance to show that $g_2=0$ also obtains for
$\Delta S=1$ hyperon beta decay, but in the real world this condition
will be violated.  A simple quark model calculation gives 
$g_2/g_1\sim -0.2$\cite{dh2} but other calculations, such as 
a recent QCD sum rule 
estimate\cite{shi} suggest a larger number---$g_2/g_1\sim -0.5$.
In any case good hyperon beta decay data---with rates {\it and} 
correlations---will be needed in order to extract the 
size of such effects since
with only decay rate information there exists a strong correlation 
between the couplings $g_1$ and $g_2$.

Recently the simultaneous SU(3) and $1/N_c$ expansion methods so successful
in understanding the hyperon masses have also been applied to the problem of
hyperon beta decay\cite{mjm}.  In this case, inclusion of symmetry breaking
terms improves the quality of fit by reducing the value $\chi^2=62.3$ 
for 23 degrees of freedom which obtains using simple SU(3) symmetry to
$\chi^2=39.2$ for 19 degrees of freedom.  However, the analysis also 
makes a prediction
that the SU(3) prediction $g_1/f_1=1.26$ for the axial coupling in 
$\Xi^0\rightarrow\Sigma^+e^-\bar{\nu}_e$ 
beta decay should be substantially reduced to
lie in the range $g_1/f_1\simeq 1.02-1.07$.  A quark model analysis by
Ratcliffe predicts a smaller reduction---$g_1/f_1\simeq 1.15$\cite{ra1}.  
The KTeV collaboration has recently
announced a preliminary number for this quantity which seems to support the
simple SU(3) analysis\cite{kt}, so clearly there is much more to be written 
on this subject.
     
\subsection{Hyperon Polarizabilities}

Since our final subject---hyperon polarizability---is not so 
familiar to many physicists, it is useful to give a bit of motivation.  
The idea goes back to simple classical
physics.  Parity and time reversal invariance, of course, forbid the existence
of a permanent electric dipole moment for an elementary particle.  
However, consider the application of a 
uniform electric field to a charged system.
Then the positive constituents making up the substructure will move in one
direction and negative charges in the other---{\it i.e.} a
charge separation will be generated leading to an {\it induced} electric 
dipole moment.  The size of
the edm will, for weak fields, be proportional to the strength of the 
applied field and
the constant of proportionality between the applied field and the induced
dipole moment is the electric polarizability $\alpha_E$
\begin{equation}
\vec{p}=4\pi\alpha_E\vec{E}
\end{equation}
The interaction of this dipole moment with the field leads to an
interaction energy
\begin{equation}
U=-{1\over 2}\vec{p}\cdot\vec{E}=-{1\over 2}4\pi\alpha_E\vec{E}^2,
\end{equation}
where the ``extra'' factor of $1\over 2$ compared to elementary physics
result is due to the feature that the dipole moment is {\it self-induced}.
Similarly in the presence of an applied magnetizing field $\vec{H}$ there will
be generated an induced magnetic dipole moment
\begin{equation}
\vec{\mu}=4\pi\beta_M\vec{H}
\end{equation}
with interaction energy
\begin{equation}
U=-{1\over 2}\vec{\mu}\cdot\vec{H}=-{1\over 2}4\pi\beta_M\vec{H}^2.
\end{equation}
One difference here is that while the electric polarizability is required on
physical grounds to be positive, the magnetic polarizability can have both
paramagnetic (from magnetic moments of the substructure constituents) and 
diamagnetic (from motion of such constituents) components. 
For wavelengths large compared to the size of the system, the effective
Hamiltonian describing the interaction of a system of charge $e$ and 
mass $m$ with an electromagnetic
field is, of course, given by 
\begin{equation}
H^{(0)}={(\vec{p}-e\vec{A})^2\over 2m}+e\phi,
\end{equation}
and the Compton scattering cross section has the simple Thomson form
\begin{equation}
{d\sigma\over d\Omega}=\left({\alpha_{em}\over m}\right)^2\left({\omega'\over
\omega}\right)^2{1\over 2}(1+\cos^2\theta),
\end{equation}
where $\alpha_{em}$ is the fine structure constant and $\omega,\omega'$ 
are the initial, final photon energies respectively.
As the energy increases, however, so does the resolution and
one must also take into account polarizability
effects, whereby the effective Hamiltonian becomes
\begin{equation}
H_{\rm eff}=H^{(0)}-{1\over 2}4\pi(\alpha_E\vec{E}^2+\beta_M\vec{H}^2).
\end{equation}
The Compton scattering cross section from such a system (taken, for simplicity,
to be spinless) is then 
\begin{eqnarray}
{d\sigma\over d\Omega}&=&\left({\alpha_{em}\over m}\right)^2\left({\omega'\over
\omega}\right)^2\left[{1\over 2}
(1+\cos^2\theta)\right.\nonumber\\
&-&\left.{m\omega\omega'\over \alpha_{em}}\left({1\over
2}(\alpha_E+\beta_M)(1+\cos\theta)^2\right.\right.\nonumber\\
&+&\left.\left.{1\over 2}(\alpha_E-\beta_M)
(1-\cos\theta)^2]\right)\right]\label{eq:sss}
\end{eqnarray}
and it is clear from Eq. \ref{eq:sss}
that from careful measurement of the differential scattering cross section,
extraction of these structure dependent polarizability terms is possible
provided 
\begin{itemize}
\item [i)] that the energy is large enough that these terms are 
significant compared to the
leading Thomson piece and 
\item [ii)] that the energy is not so large that higher order
corrections become important.  
\end{itemize}
In this fashion the measurement of electric and
magnetic polarizabilities for the proton has recently been accomplished 
at SAL and at MAMI using
photons in
the energy range 50 MeV  $<\omega <$ 100 MeV, yielding\cite{PPol}
\begin{eqnarray}
\alpha_E^p&=&(12.1\pm 0.8\pm 0.5)\times 10^{-4}\; {\rm fm}^3\nonumber\\
\beta_M^p&=&(2.1\mp 0.8\mp 0.5)\times 10^{-4}\; {\rm fm}^3. \label{abexp}
\end{eqnarray}
(Results for the neutron extracted from $n-Pb$ scattering cross
section
measurements have been reported\cite{npol}, 
but have been questioned\cite{ques}.
Extraction via studies using a deuterium target may be possible
in the future\cite{bean}.)
Note that in
practice one generally exploits the strictures of causality and unitarity as
manifested
in the validity of the forward scattering dispersion relation, which yields the
Baldin sum rule\cite{bgm}
\begin{eqnarray}
\alpha_E^{p,n}&+&\beta_M^{p,n}={1\over 2\pi^2}\int_0^\infty{d\omega\over
\omega^2}\sigma_{\rm tot}^{p,n}\nonumber\\
&=&\left\{
\begin{array}{ll}(13.69\pm 0.14)\times 10^{-4}{\rm fm}^3& {\rm proton}\\
                 (14.40\pm 0.66)\times 10^{-4}{\rm fm}^3& {\rm neutron}
\end{array}\right.\nonumber\\
\quad
\end{eqnarray}
as a rather precise constraint because of the small uncertainty associated 
with the photoabsorption cross section $\sigma_{\rm tot}^p$.

As to the meaning of such results we can compare with the corresponding 
calculation of the electric polarizability of the hydrogen atom, 
which yields\cite{merz}
\begin{equation}
\alpha_E^H={9\over 2}a_0^2\quad{\rm vs.}\quad\alpha_E^p\sim
10^{-3}<r_p^2>^{3\over 2}
\end{equation}
where $a_0$ is the Bohr radius.  Thus the polarizability of the hydrogen
atom is of order the atomic volume while that of the proton is 
only a thousandth of its volume, indicating that the proton is 
much more strongly bound.

On the theoretical side, the electric and magnetic polarizability of the 
nucleon have been evaluated in heavy baryon chiral perturbation theory.  
At one loop---${\cal O}(p^3)$---one finds that the theoretically 
predicted numbers follow strictly from chiral symmetry and are in spectacular
agreement with the experimentally measured values\cite{bkm}
\begin{equation}
\alpha_E^p=10\beta_M^p={5\alpha g_A^2\over 96\pi F_\pi^2 m_\pi}=12.2
\times 10^{-4}\,{\rm fm}^3
\end{equation}
The ${\cal O}(p^4)$ terms have also been evaluated.  At this order
uncertainties due to unknown counterterms develop, but the basic agreement
remains, within errors\cite{bkm1}.

The relevance of this work to hyperon physics is that the
polarizability of a hyperon can also be measured using Compton
scattering, via the reaction $B+Z\rightarrow B+Z+\gamma$ extrapolated to
the photon pole---{\it i.e.} the Primakoff effect.   Of course, this is
only feasible for charged hyperons---$\Sigma^\pm,\Xi^-$, and the 
size of such polarizabilities predicted
theoretically via SU(3) chiral perturbative techniques are  
somewhat smaller than that of the proton\cite{meis}
\begin{equation}
\alpha_E^{\Sigma^+}\sim9.4\times 10^{-4}\,{\rm fm}^3,\qquad
\alpha_E^{\Xi^-}\sim2.1\times 10^{-4}\,{\rm fm}^3
\end{equation}
but their measurement would be of great interest and would provide a 
new illumination of hyperon structure.

\section{Summary}  
I conclude by noting that, although the first hyperon was discovered
more then half a century ago and much work has been done since, 
the study of hyperons remains an interesting and challenging
field.  As I have tried to indicate above, many questions still exist
as to their weak and electromagnetic interaction properties,
and I suspect that such particles will remain choice targets for particle
hunters well into the next century.

\begin{center}
{\bf Acknowledgement}
\end{center}

It is a pleasure to acknowledge helpful comments by E. Henley and R. Lebed
as well as the hospitality of CTP and LNS at MIT and of 
INT and the Department of Physics at the University of Washington 
where this paper was written.

\end{document}